\documentclass[5p,times]{elsarticle}

\usepackage{amssymb}

\usepackage{tabularx}
\usepackage{listings}
\usepackage{xcolor}
\usepackage{subcaption}
\usepackage{float}
\usepackage{hyperref}

\definecolor{codegreen}{rgb}{0,0.6,0}
\definecolor{codegray}{rgb}{0.5,0.5,0.5}
\definecolor{codepurple}{rgb}{0.58,0,0.82}
\definecolor{backcolour}{rgb}{1.0,1.0,1.0}
\definecolor{java_comment_green}{rgb}{0.301,0.51,0.4}
\definecolor{gray}{rgb}{0.4,0.4,0.4}
\definecolor{darkblue}{rgb}{0.0,0.0,0.6}
\definecolor{cyan}{rgb}{0.0,0.6,0.6}

\definecolor{java_string_color}{rgb}{0.784,0.553,0.458}
\definecolor{attribute_color}{rgb}{0.365, 0.513, 0.588}

\definecolor{maroon}{rgb}{0.5,0,0}
\definecolor{darkgreen}{rgb}{0,0.5,0}
\lstdefinelanguage{XML}
{
  basicstyle=\ttfamily\footnotesize,
  morestring=[s]{"}{"},
  morecomment=[s]{?}{?},
  morecomment=[s]{!--}{--},
  commentstyle=\color{darkgreen},
  moredelim=[s][\color{black}]{>}{<},
  moredelim=[s][\color{attribute_color}]{\ }{=},
  stringstyle=\color{java_string_color},
  identifierstyle=\color{darkblue}
}

\definecolor{dkgreen}{rgb}{0,0.6,0}
\definecolor{gray}{rgb}{0.5,0.5,0.5}
\definecolor{mauve}{rgb}{0.58,0,0.82}

\lstdefinestyle{Python}{
    language        = Python,
    basicstyle      = \ttfamily,
    keywordstyle    = \color{blue},
    keywordstyle    = [2] \color{teal}, 
    stringstyle     = \color{green},
    commentstyle    = \color{red}\ttfamily
}

\newcolumntype{Y}{>{\centering\arraybackslash}X}
\newcolumntype{C}[1]{>{\centering\arraybackslash}p{#1}}

\journal{Future Generation Computer Systems}

\begin{document}

\begin{frontmatter}

\title{CLAID: Closing the Loop on AI \& Data Collection - A Cross-Platform Transparent Computing Middleware Framework for Smart Edge-Cloud and Digital Biomarker Applications}

\author[inst1]{Patrick Langer\corref{cor1}}
\author[inst1,inst2]{Elgar Fleisch}
\author[inst1]{Filipe Barata}

\affiliation[inst1]{organization={ETH Zurich},
            addressline={Centre for Digital Health Interventions}, 
            city={Zurich},
            country={Switzerland}}

\affiliation[inst2]{organization={University of St. Gallen},
            addressline={Centre for Digital Health Interventions}, 
            city={St. Gallen},
            country={Switzerland}}
\cortext[cor1]{Corresponding author. E-mail address: \href{mailto:planger@ethz.ch}{planger@ethz.ch}}

\begin{abstract}
The increasing number of edge devices with enhanced sensing capabilities, such as smartphones, wearables, and IoT devices equipped with sensors, holds the potential for innovative smart-edge applications in healthcare. These devices generate vast amounts of multimodal data, enabling the implementation of digital biomarkers which can be leveraged by machine learning solutions to derive insights, predict health risks, and allow personalized interventions. Training these models requires collecting data from edge devices and aggregating it in the cloud. To validate and verify those models, it is essential to utilize them in real-world scenarios and subject them to testing using data from diverse cohorts. Since some models are too computationally expensive to be run on edge devices directly, a collaborative framework between the edge and cloud becomes necessary. In this paper, we present CLAID, an open-source cross-platform middleware framework based on transparent computing compatible with Android, iOS, WearOS, Linux, macOS, and Windows. CLAID enables logical integration of devices running different operating systems into an edge-cloud system, facilitating communication and offloading between them, with bindings available in different programming languages. We provide Modules for data collection from various sensors as well as for the deployment of machine-learning models. Furthermore, we propose a novel methodology, \textit{ML-Model in the Loop} for verifying deployed machine learning models, which helps to analyze problems that may occur during the migration of models from cloud to edge devices. We verify our framework in three different experiments and achieve 100\% sampling coverage for data collection across different sensors as well as an equal performance of a cough detection model deployed on both Android and iOS devices. Additionally, we compare the memory and battery consumption of our framework across the two mobile operating systems.
\end{abstract}


\end{frontmatter}

\section{Introduction}
\label{sec:introduction}

The proliferation of portable edge devices including wearable sensors, mobile phones, and Internet-of-Things (IoT) devices with sensing capabilities, such as implantables, or digestibles~\cite{iot_sensors_review}, holds great promise in meeting the health monitoring requirements of the global aging population and patients with chronic diseases~\cite{ncd_costs_switzerland,ncd_financial_burden,ncd_demographic_transition, nature_digital_biomarkers}.  
By providing individuals with near real-time data on their health status, edge devices can enable a more proactive approach to health management including early detection of diseases, predicting treatment outcomes, and promoting personalized interventions\cite{increasing_use_of_smartphones_mhealth, review_application_of_smarthpones_in_disease_monitoring}. 
Examples of near real-time data may include physiological data, such as photoplethysmography, as well as behavioral data like actigraphy~\cite{nature_digital_biomarkers, review_application_of_smarthpones_in_disease_monitoring}. 
To be useful to physicians and patients, however, unstructured sensor data collected by these devices must be processed and aggregated to create digital biomarkers often using machine learning. Digital biomarkers are defined as physiological and behavioral data that are collected and measured by means of edge devices that explain, influence, or predict health-related outcomes~\cite{nature_digital_biomarkers, sim2019mobile, digital_biomarker, digital_phenotype}.  
Machine-learning approaches are essential for the development of digital biomarkers, as they help derive meaningful insights from the vast amounts of behavioral and physiological data generated by edge devices. By applying these algorithms, hidden patterns in the collected data can be revealed, leading to a deeper understanding of an individual's health status and the ability to assess it accurately~\cite{ml_in_healthcare}. Additionally, machine learning algorithms can be used to perform medical functions, such as treatment recommendations, interventions~\cite{nature_digital_biomarkers}, or provide new sensing capabilities~(e.g., automated cough detection from audio data~\cite{filipe_cough_detection}).

In recent years, there has been a growing recognition of the need for structured and reusable software frameworks to support the development of digital biomarkers in terms of data collection~\cite{aware,sensus,radar_base} and data analysis~\cite{digital_biomarker_discovery_pipeline, flirt}. The primary goal of such software frameworks is to minimize the complexity and thus the 
effort associated with managing various devices and sensors during development. To accomplish this goal, these frameworks offer reusable components, such as for data acquisition, upload, and analysis to streamline and simplify the overall process of developing digital biomarkers. In accordance with prior work~\cite{nature_digital_biomarkers}, we derive the requirements that a software framework for the development and evaluation of digital biomarkers must satisfy: 
\begin{enumerate}
    \item \textbf{modularity} is necessary to enable digital biomarkers to be composed of interoperable software and hardware components. In this context, modularity refers to the flexible implementation of components for sensor data collection, machine learning models, or computations. These components have to be easily reconfigurable with little-to-no coding, to enable the transition from data collection to data analysis and to facilitate their reuse in different scenarios. Further, they have to be deployable across different hardware devices like smartphones or smartwatches.
    \item \textbf{measurements} are required for digital biomarkers. The measurements are captured by sensors and allow the creation of multimodal datasets during medical studies, which are required for the training of machine learning models for digital biomarkers. The data is commonly aggregated in the cloud, i.e., multiple servers equipped with ample storage and computing capacities. To collect data, therefore, edge devices need to be equipped with software for capturing, processing, and uploading data, while cloud servers need databases for storage. Ideally, data acquisition from sensors can be set up without programming, by simple configuration files or user interfaces, given that support for a sensor is integrated in the framework. Adding support for further sensors, like external wearables, should be facilitated by the framework.

    \item \textbf{verification} and \textbf{validation} are essential for ensuring the performance and applicability of digital biomarker models across diverse patient populations and contexts. Verification analytically ensures that the output of a digital biomarker model remains accurate across diverse hardware and computing capabilities. Validation empirically confirms that the digital biomarker model reliably predicts or monitors health outcomes as intended across different devices and patients, for example by testing them in clinical settings and with different cohorts. Deploying digital biomarker models is a prerequisite for verification and validation, as it enables the assessment of data and predictions based on continuous measurements by edge devices under real-life conditions. Since edge devices have only limited computing and memory capabilities, deploying models on those devices might not always be feasible. Some models might be too computationally or memory intensive~\cite{smartphones_ml_resouce_limitation} to be directly deployed on edge devices and therefore have to be deployed in the cloud. A framework for the evaluation of digital biomarkers must therefore enable the deployment of machine learning models on edge devices, cloud devices, or both where necessary.
\end{enumerate}

To date, however, no framework addressing all three requirements for developing and evaluating digital biomarkers has emerged. Prior work only partially addresses these requirements and focuses on individual aspects. For example, existing solutions like Pogo~\cite{pogo_middleware} and USense~\cite{usense} focus on modularity but have not been developed for medical use cases and lack support for external sensors, such as wearables, and integration of machine learning models. Sensing frameworks such as Sensus~\cite{sensus}, RADAR-BASE~\cite{radar_base}, and CAMS~\cite{carp_cams} target measurements, but do not offer the necessary modularity for incorporating machine learning models. Lastly, frameworks such as MobiCOP~\cite{mobicop, mobicop_iot} and MTC~\cite{mtc_new} can enable machine learning model deployment by leveraging edge and cloud resources to offload computations. 
Nevertheless, these frameworks have not been tailored for medical scenarios and lack the flexibility seen in frameworks like Pogo. Additionally, they do not offer the capability to integrate sensor-based measurements.

To contribute towards a software framework for the development and evaluation of digital biomarkers addressing all three requirements, we have designed, implemented, and evaluated the CLAID~(Closing the Loop on AI and Data Collection) framework. CLAID is a cross-platform, cross-language research and development middleware framework based on transparent computing, offering capabilities for data collection and machine learning deployment. 
To achieve the required modularity, we apply several software design principles found in middleware, transparent computing, and offloading. As in middleware, which can be considered as an abstraction layer between the operating system and applications~\cite{middleware_review_paper, middleware_iot}, we provide functionality for implementing flexible and loosely coupled components which we refer to as Modules. We offer concrete implementations of Modules for sensing and model deployment, targeting data collection and analysis. To provide flexible communication between these Modules, we use the network computing paradigm of transparent computing, which offers abstractions for communication between heterogeneous devices and operating systems~\cite{edge_end_cloud_transparent_computing}. Using transparent computing, network connections and data transfers remain imperceptible or \textit{transparent} to the Modules, hence they can communicate seamlessly even across network boundaries. The transparent communication between individual Modules enables the integration of edge and cloud devices into a logical edge-cloud system, which allows the distribution of Modules throughout the system as required. This distribution enables offloading~\cite{offloading_algorithms, static_and_dynamic_offloading, offloading_fog_iot}, a process that allows transferring computational tasks, such as inference of machine learning models, from edge devices to the cloud where necessary. The key contributions of our work, addressing specific requirements, encompass the following points:

\begin{enumerate}
    \item We provide a scalable, flexible, and self-contained cross-platform \textbf{middleware} implementation enabling the creation of loosely coupled Modules for example for data collection and model deployment, which is compatible with different programming languages and operating systems, targeting the modularity requirement~(1).
    \item We provide existing Modules for data collection and an API to facilitate the integration of new Modules, targeting the requirement for measurements~(2). We also provide existing Modules for background recording, data storage, and upload, that help mitigate common pitfalls encountered when collecting data on mobile devices. We enable the combination and reconfiguration of existing Modules as required from configuration files.
    \item We integrate mechanisms of \textbf{transparent computing} into the middleware, allowing the combination of edge and cloud devices into a logical edge-cloud system. Further, we provide \textbf{offloading} capabilities which allow the distribution of Modules across this edge-cloud system. These capabilities allow the deployment of machine learning models on both, edge and cloud devices to fulfill the third requirement, the verification and validation of developed models in real-world scenarios~(3).
    \item We evaluate our framework in three separate experiments. With the first experiment, we verify stable data collection with different sensors over 24 hours, using sampling coverage as a metric. With the second experiment, we propose a novel methodology for the verification of deployed machine-learning models, which we term~\textit{ML-Model in the Loop}. With the third experiment, we assess the memory and battery utilization of our framework. 
\end{enumerate}
CLAID's primary focus is facilitating the development of digital biomarkers, achieved by harnessing mobile devices like smartphones, wearables, and Bluetooth peripherals to collect datasets for training machine learning-based digital biomarkers. Yet, its potential extends beyond this specific domain, offering broader applicability.  We consider it to be useful for the development of smart edge or edge-cloud applications in general. As a contribution to the research community, we release CLAID as a fully open-source framework and provide extensive documentation, tutorials, and example applications via a dedicated website~(cf. Table~\ref{tab:introduction:code_and_website}).
\begin{table}[h!t]
\caption{Code Availability}
\label{tab:introduction:code_and_website}
\centering
\renewcommand*{\arraystretch}{1.4}
\begin{tabularx}{\columnwidth}{l|X}
\hline
\textbf{Implementation} & C++, Java, Python, Objective-C\\
\textbf{Documentation} & \url{https://www.claid.ethz.ch}\\
\textbf{Code repository} & \url{https://github.com/ADAMMA-CDHI-ETH-Zurich/CLAID}\\
\hline
\end{tabularx}
\end{table}

The remainder of this work is organized as follows: Section~\ref{sec:related_work} provides an overview of related work, such as transparent computing, middleware, and mobile sensing frameworks. Section~\ref{sec:methods} provides a system- and architectural overview of CLAID. Section~\ref{sec:experiments} explains the experiments conducted for evaluation, with results presented in section~\ref{sec:results}. Lastly, sections~\ref{sec:discussion} and~\ref{sec:outlook} conclude with a discussion and future outlook.
\section{Related Work}
\label{sec:related_work}

In this section, we highlight previous research related to CLAID in terms of middleware frameworks, transparent computing, offloading, and mobile sensing frameworks.  In Table~\ref{tab:related_frameworks}, we provide a comparison of relevant frameworks and implementations related to CLAID.

\begin{table*}[h!t]
  \caption{Comparison of related frameworks and their features.}
  \label{tab:related_frameworks}
  \begin{tabularx}{\textwidth}{C{2.5cm}p{2.2cm}C{3cm}YYYYYC{1.9cm}}
    Framework & Type & OS &Internal \mbox{Sensors} & External \mbox{Sensors} & Extension API & ML \mbox{Models} & Offloa- ding & Code-free reconfiguration \\\hline
    CLAID (ours) & Transparent Computing Middleware &Android, WearOS, iOS, Linux, macOS, Windows & \checkmark & \checkmark & \checkmark & \checkmark & \checkmark & \checkmark \\ \hline 
    USense~\cite{usense} & Mobile \mbox{Middleware} &Android &  \checkmark & \texttimes & \texttimes & \texttimes & \texttimes & \texttimes \\
    Pogo~\cite{pogo_middleware} & Mobile \mbox{Middleware} & Android, Linux, macOS, Windows & \checkmark & \texttimes & \checkmark & \texttimes & \texttimes & \texttimes \\
    MTC~\cite{mtc_new} & Transparent Computing Framework & Android, Linux  &  \texttimes & \texttimes & \checkmark & \texttimes & \checkmark & \texttimes \\
    MobiCOP~\cite{mobicop,mobicop_iot} & Mobile \mbox{Middleware} & Android &  \texttimes & \checkmark & \checkmark & \texttimes & \checkmark & \texttimes \\
    
    \hline 
    AWARE~\cite{aware} & Mobile Sensing Framework & Android, iOS & \checkmark & \checkmark & \checkmark & \texttimes    & \texttimes & \checkmark \\  
    Beiwe~\cite{beiwe} & Mobile Sensing Framework & Android, iOS & \checkmark & \texttimes & \texttimes & \texttimes  & \texttimes & \checkmark \\  
    CAMS \cite{carp_cams} & Mobile Sensing Framework & Android, iOS & \checkmark & \checkmark & \checkmark & \texttimes    & \texttimes & \texttimes \\  
    mCerebrum \cite{mcerebrum} & Mobile Sensing Framework & Android & \checkmark & \texttimes & \texttimes & \texttimes    & \texttimes & \texttimes \\  
    RADAR-base~\cite{radar_base} & Mobile Sensing Framework &  Android  & \checkmark &\checkmark & \checkmark & \texttimes  & \texttimes & \texttimes \\ 
    Sensus~\cite{sensus} & Mobile Sensing Framework & Android, iOS & \checkmark & \checkmark & \texttimes & \texttimes  & \texttimes & \texttimes \\ \hline
    
    \end{tabularx}
    \caption*{"Internal sensors" refers to sensors built into smartphones, such as GPS or microphones, while "external sensors" refers to those integrated with wearables or other devices. An "Extension API" allows the integration of new functionalities. "ML Algorithms" indicates the framework's capabilities for deploying machine learning algorithms. "Offloading" refers to the ability to execute functions or Modules across different devices~(e.g. smartphone and server) seamlessly. Lastly, "Code-free reconfiguration" enables the loading and configuration of different components without the need for additional coding, for example via configuration files.}
\end{table*}
\subsection{Middleware Frameworks}
\label{sec:middleware_frameworks}
Middleware can be considered as an abstraction layer that sits between the operating system and software applications. It is designed to manage hardware heterogeneity, improve software application quality, simplify software design, and reduce development costs~\cite{elkady2012robotics}. Middleware provides functionalities for the implementation of loosely coupled components, that can communicate, exchange data and perform tasks. In recent years, middleware frameworks have been the subject of increasing research in fields such as distributed systems, the Internet of Things~(IoT), cloud computing, and robotics~\cite{middleware_review_paper, middleware_iot, middleware_cloud_computing, ros}. Especially the field of robotics has seen a proliferation of advanced middleware frameworks and architectures~\cite{middleware_review_paper}. Among the most widely used robotic middleware is the Robot Operating System~(ROS)~\cite{ros}, which has recently been succeeded by ROS2~\cite{ros2} and has become the de-facto standard for robotic applications in industry and research. ROS provides many packages and features for sensing tasks, perception, and control~\cite{ros2}.  

Some middleware implementations are specifically designed to be deployed on mobile devices such as smartphones and can offer functionalities for integrating sensors, context-aware applications, and synchronization with the cloud. The Pogo middleware~\cite{pogo_middleware}, is designed to assist researchers in setting up data collection from smartphones and aims to enable real-world deployment and evaluation of developed algorithms. It provides an API enabling to add custom functionality. The Pogo middleware was published in 2012 and the source code is not publicly available. USense, another mobile middleware, was published in 2014~\cite{usense}. USense is focused on community-driven sensing tasks and incorporates concepts such as application awareness, user awareness, and situation awareness. It aims to identify appropriate situations and collect data when an opportunity arises. USense, however, only provides a limited set of sensors, which cannot necessarily be queried at any time. Instead, the hardcoded settings of the middleware determine when to record data. Consequently, it cannot be considered a modular middleware. Both, Pogo and USense, do not offer support for machine learning models.

In general, the existing mobile middleware solutions often do not offer the same type of extensibility and features found in robotic middleware frameworks. 
Robotic middleware frameworks often feature a packaging system that allows them to include new features, with frameworks like ROS providing numerous packages containing components for data collection and analysis. Robotic middleware, however, cannot readily be used for mobile application scenarios. They are primarily developed for desktop or embedded operating systems such as Linux, Windows, and macOS. As a result, they often depend on third-party libraries and frameworks, such as boost~\cite{boost}, which may only be available for those desktop operating systems. Furthermore, ROS-Mobile~\cite{ros_mobile}, an implementation that can be used to connect Android applications to instances of ROS and control them to some extent, does not enable running ROS itself. To date, there is no official support for Android or iOS for either ROS or ROS2. To the best of our knowledge, there are currently no middleware frameworks publicly available that allow the integration of mobile, IoT, and cloud devices. This can be explained by the inherent complexity of managing heterogeneous devices and operating systems.

\subsection{Transparent Computing}
The communication between heterogeneous edge and cloud devices can be enabled by transparent computing~(TC). Transparent computing is a network computing paradigm aiming to provide abstractions to manage the heterogeneity of devices, operating systems, and resources~\cite{edge_end_cloud_transparent_computing}. 
According to Ren et al, "transparent computing enables devices to choose services on demand via networks"~\cite{edge_end_cloud_transparent_computing}. Those services, among others, can entail data storage as well as computational services. For example, a service running on a smartphone could request to use a service to store large amounts of data. If such a service is provided, e.g., by a connected cloud device, data can be communicated between these two services. The service running on the smartphone does not need to know, where the storage service is running. The data will be forwarded between the two services by the transparent computing mechanism automatically. Hence, from the viewpoint of the services, transfer of data or network communication is hidden or \textit{transparent}. A concrete implementation of a transparent computing approach is provided by Zhou et al.~\cite{mtc_new}. They propose a mobile transparent computing framework "Mobile Transparent Computing" (MTC), unifying communication across edge and cloud devices. This framework allows computations to run across local, nearby, or remote devices. The distributions of those computations are referred to as offloading.

\subsection{Offloading}
Offloading is a technique that can be used to distribute tasks and computations across devices of different capabilities. Different types of offloading mechanisms exist. Offloading can be done static, or dynamic~\cite{static_and_dynamic_offloading}. Static means that the distribution of computations is determined only once, for example during compile time or start of the application. This can be done, for example, by analyzing available resources during startup, or simply manually as configured by a user or developer~\cite{static_and_dynamic_offloading}. Dynamic offloading, on the other hand, allows one to adapt more flexibly and leverage computing resources based on availability by using load balancing or energy balancing strategies~\cite{mobicop}. Furthermore, different granularities of offloading can be distinguished. For example, methods, classes, threads, or whole applications can be offloaded. Offloading can be implemented independently of transparent computing. For instance, Benedetto et al. propose the MobiCOP framework for offloading between Android devices~\cite{mobicop}. MobiCOP does not use principles of TC, but follows a fixed client-server structure, with smartphones as clients and cloud devices as servers. It specifically relies on the Android Operating System, also for cloud devices. With MobiCOP-IoT~\cite{mobicop_iot}, support for devices running Android Things is available as well, hence it is possible to use IoT devices running this operating system for data collection. While offloading does not necessarily depend on transparent computing, transparent computing can facilitate its implementation, and enable offloading between heterogeneous operating systems.

\subsection{Mobile Sensing Frameworks}
\label{sec:sensing_frameworks}
Mobile sensing frameworks align with the trend of cloud-based analysis of the previous decade~\cite{edge_end_cloud_transparent_computing}. They work unidirectionally, transferring data from mobile devices to the cloud for subsequent analysis. 
While not as flexible as middleware frameworks, they can offer comprehensive solutions for data collection from mobile devices, and different frameworks have emerged over the past decade.  AWARE~\cite{aware}, released in 2010, was one of the first open-source frameworks that aimed to provide a complete and configurable solution for data collection from mobile devices~\cite{research_frameworks_survey_paper}. AWARE mainly focuses on internal smartphone sensors like GPS or audio as well as questionnaires, but it features a plugin API that allows the inclusion of new internal and external sensors~\cite{aware}. It comes with an Android and iOS version, although both are maintained separately and do not share a common code base. This leads to differences in the API on both platforms. Additionally, the AWARE applications depend on the AWARE web server in order to upload data. Using a custom server setup is not easily possible.
Another mobile sensing framework is the Beiwe platform~\cite{beiwe}, which comes with two different versions for Android and iOS as well and only supports a limited amount of internal phone sensors~(such as GPS and phone logs). It does not provide an API to include external sensors. Furthermore, data collection in Beiwe can be configured from a dashboard running on a web server. RADAR-base targets Android devices\footnote{an iOS version exists with limited functionality; only allows to record location data without uploading it to a server} and focuses especially on security aspects. It employs different authentication measures and strategies to ensure safe access to recorded data for analysis and visualization through web interfaces~\cite{radar_base}. RADAR-base generally supports both internal and external sensors, however, access to integration of external sensors is kept as a closed-source feature~\cite{radar_base, carp_cams}. The mCerebrum framework, also exclusively available for Android, aims to achieve high-throughput data collection by prioritizing high-sampling rate recording~\cite{mcerebrum}. In contrast to the aforementioned frameworks, Sensus~\cite{sensus} and CAMS~\cite{carp_cams} pursue a \textit{crossplatform} approach, which means they use common programming languages (C\# via Xamarin~\cite{xamarin} for Sensus and Dart via Flutter~\cite{flutter} for CAMS) that can run on both Android and iOS, instead of maintaining separate code bases. On one hand, this allows for a high level of flexibility and the ability to target multiple devices and platforms. On the other hand, since C\# and Dart are not native languages for Android and iOS, these frameworks have a limitation in that developers cannot directly access hardware sensors on those platforms. Instead, they have to rely on additional packages or manually develop them to realize sensor integration using the native languages of Android and iOS and provide an interface to be used in C\# or Dart. Those packages, however, often can not offer the same functionalities as found in the native APIs and are updated less frequently, as mentioned by the authors of CAMS as well.

\section{Methodology}
\label{sec:methods}
In this section, we introduce CLAID, a transparent computing middleware framework stemming from our digital biomarker research. With CLAID, we seek to realize a framework that fulfills the requirements for the development of digital biomarkers as derived in section~\ref{sec:introduction}. To meet these requirements, we apply software principles found in middleware, transparent computing, offloading, and mobile sensing, which we describe in section~\ref{sec:claid_framework}. We evaluate our framework against the requirements using three different experiments, as described in section~\ref{sec:experiments}.

\subsection{CLAID framework implementation}
\label{sec:claid_framework}
\begin{figure}[h!t]
  \centering
  \includegraphics[width=1.0\columnwidth]{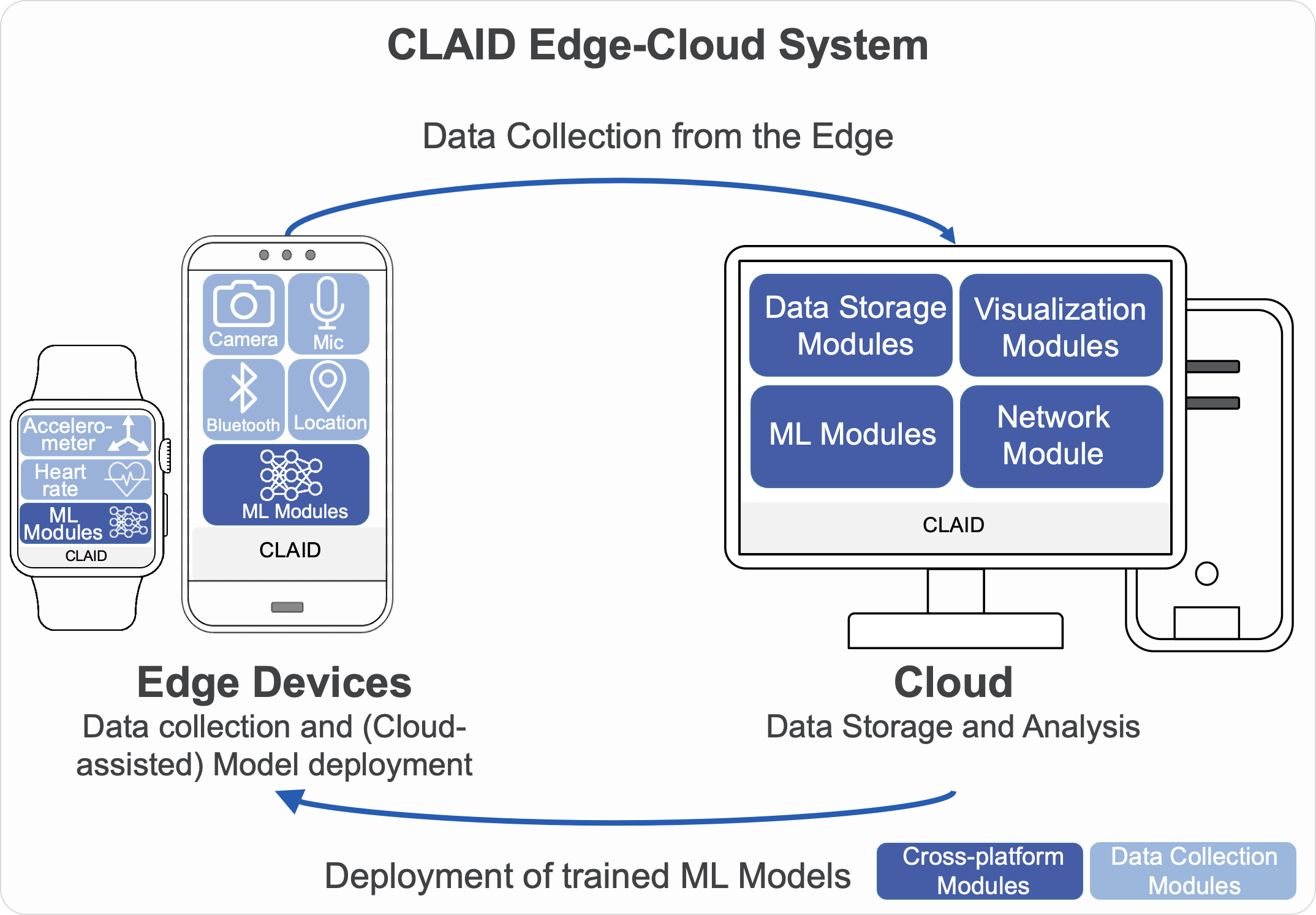}
  \caption{
  Researchers and developers can create CLAID applications using Modules designed for sensor integration, machine learning deployment, algorithms, or visualizations. We offer platform-specific Sensor Modules deployable on mobile operating systems (Android, iOS, WearOS), along with cross-platform modules that work on both mobile and desktop (Windows, Linux, macOS) operating systems. The cross-platform Modules enable the deployment of trained models either on the Server or directly on the Edge device. Modules can communicate seamlessly over a network connection, allowing Models to directly utilize data recorded on the Edge device in both cases.}
  \label{fig:application_overview}
\end{figure}

To effectively validate and apply digital biomarker-based applications and the treatment recommendations derived from them, we require an approach that allows us to repurpose our data collection applications to seamlessly integrate our developed models. Hence, we strive to close the loop from data collection to model deployment. Figure ~\ref{fig:application_overview} shows a high-level concept of the proposed closed-loop approach.
 Our framework allows building applications by combining different components, which we refer to as Modules. These Modules realize functionalities such as sensor integration, data storage, and deployment of machine learning models. Using CLAID, we can implement Modules in different programming languages and combine them as required using configuration files written in XML. The configuration files allow us to reconfigure and recombine Modules for different application scenarios, allowing us to switch from data collection to model deployment. 
\subsubsection{Middleware architecture}
\label{sec:claid_architecture}
\begin{figure}[h!t]
  \centering
  \includegraphics[width=1.0\linewidth]{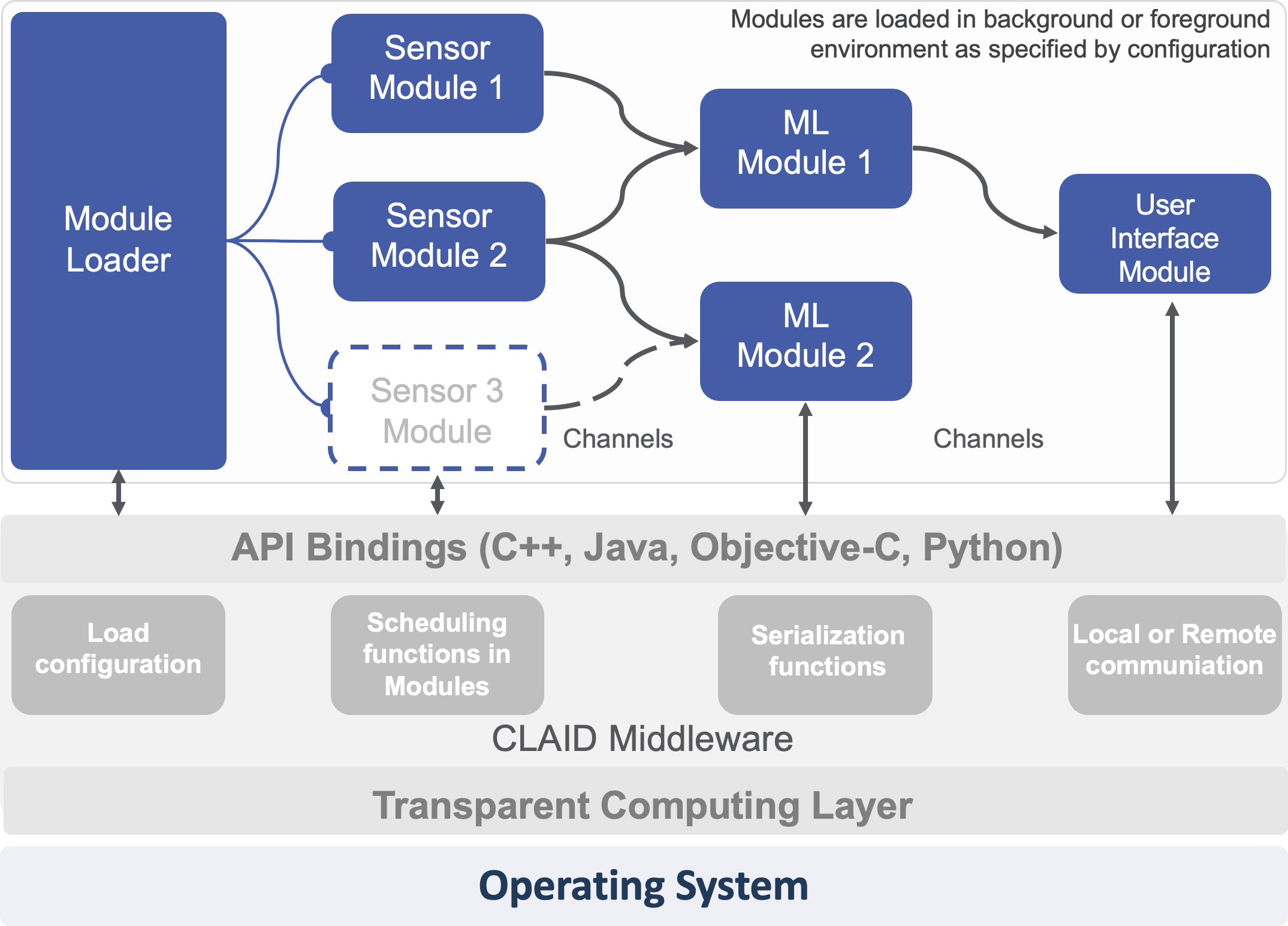}
  \caption{To build applications with CLAID, we can combine different Modules~(blue) realizing sensor integration, machine learning deployment, and more. Except for sensor Modules, the pre-existing Modules work on different mobile~(Android, iOS) as well as desktop~(Windows, Linux, macOS) operating systems. The middleware functions uniformly across all these operating systems.}
  \label{fig:middleware_architecture}
\end{figure}

Our framework is based on a middleware architecture, which offers functionalities for scheduling, configurations, serialization, and communication, which developers can use to implement Modules. Modules can communicate via channels based on the publisher-subscriber paradigm~\cite{publish_subscribe_different_implementations} commonly found in middleware frameworks like ROS~\cite{ros}. Channels are bidirectional and thus allow each Module to communicate with any other Module. Each channel is identified by a \textit{name} or \textit{topic}, which is specified when subscribing to or publishing a channel. Multiple instances of CLAID can connect to each other via a network. We implement mechanisms of transparent computing that allow Modules to communicate even across a network border. If a Module posts data to a channel that has a subscriber in a remotely connected instance, the data will be serialized and sent to the remotely running Module automatically~(see section~\ref{sec:transparent_computing}). Otherwise, communication happens only locally. Figure~\ref{fig:middleware_architecture} depicts an overview of the CLAID middleware architecture.

We base our framework on a common code base in C++, which allows for native compatibility with all targeted operating systems. Using C++, we create a common API and provide bindings to use our framework in a variety of native languages like Java and Python. This means that while the base functionality of CLAID is implemented in C++, developers can use and extend it using languages such as Java, Objective-C, and Python. An example of a Module implemented using the Java API is shown in Listing~\ref{listing:claid_module}.
\begin{figure}[h!t]
\noindent\begin{minipage}[h!t]{\columnwidth}
\begin{lstlisting}[caption=Example of a CLAID Module implemented via the JavaCLAID API,frame=tlrb,language=Java,label={listing:claid_module},captionpos=b]{Name}
class MyModule extends JavaCLAID.Module
{
    private Channel<String> receiveChannel;
    private Channel<String> sendChannel;
    void initialize()
    {
        receiveChannel = subscribe<String>(
            "DataChannel", data -> onData(data));
        
        sendChannel = publish<String>("DataChannel");
        sendChannel.post("Hello World");
        
        registerPeriodicFunction(
            500, () -> periodicFunction);  
            
        registerScheduledFunction(
            Date.everyDay(), Time(13, 37, 00),  
            () -> scheduledFunction());
    }

    // Called, when new data arrived at "DataChannel"
    void onData(String data)
    {
        System.out.println(data);
    }

    // This function will be called every 500ms.
    void periodicFunction()
    {
        channel.post("Periodic function called.");
    }

    // This function will be called every day at 13:37:00
    void scheduledFunction()
    {
        System.out.println(Time.now().toString());
    }
};
\end{lstlisting}
\end{minipage}
\end{figure}
To create API bindings for different languages, we currently employ pybind11~\cite{pybind11} and jbind11\footnote{Note, that jbind11 was specifically implemented by us to mimic the functionality of pybind11 for Java and will be released separately}~\cite{jbind11}. The API provides functionalities for the creation of Modules, communication between Modules through channels, serialization and deserialization, properties, and scheduling of functions at specific times or interval rates. We implement cross-platform Modules directly in C++ to be able to reuse them across different devices.
To collect data, however, we create platform-specific Modules using one of the native languages for each mobile operating system, such as Java or Objective-C. This approach allows us to use the native API of the mobile OS directly, ensuring maximum compatibility with the offered functionalities. To combine existing Modules to create applications, we provide a ModuleLoader that allows loading the required Modules at runtime. The ModuleLoader processes configuration files, instantiates individual Modules using their name or identifier, and assigns arbitrary properties according to the configuration file. For Android and iOS, the ModuleLoader can automatically place Modules in a foreground or background environment, for which a Service is launched if required. This flexibility enables the use of Modules in various scenarios. CLAID and our API are extensible, and we provide a package managing system that allows us to incorporate new functionality.

\subsubsection{Reconfiguration: Data Collection to Model Deployment}
\begin{figure*}[h!t]     \centering
     \begin{subfigure}{0.48\textwidth}
         \centering
         \includegraphics[width=1.0\textwidth]{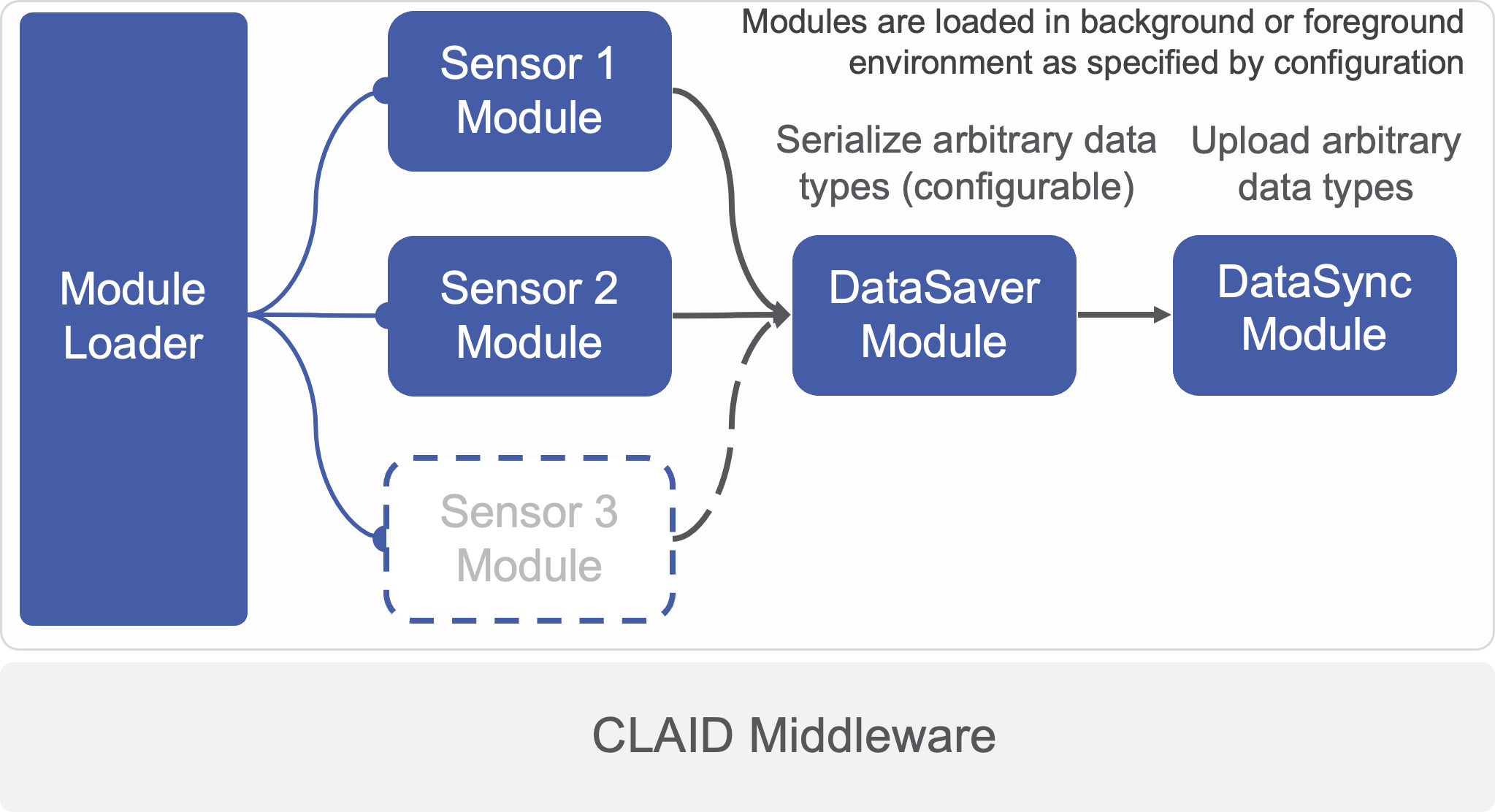}
         \caption{CLAID Modules configured for Data Collection.}
     \end{subfigure}
     \begin{subfigure}{0.48\textwidth}
         \centering
         \includegraphics[width=1.0\textwidth]{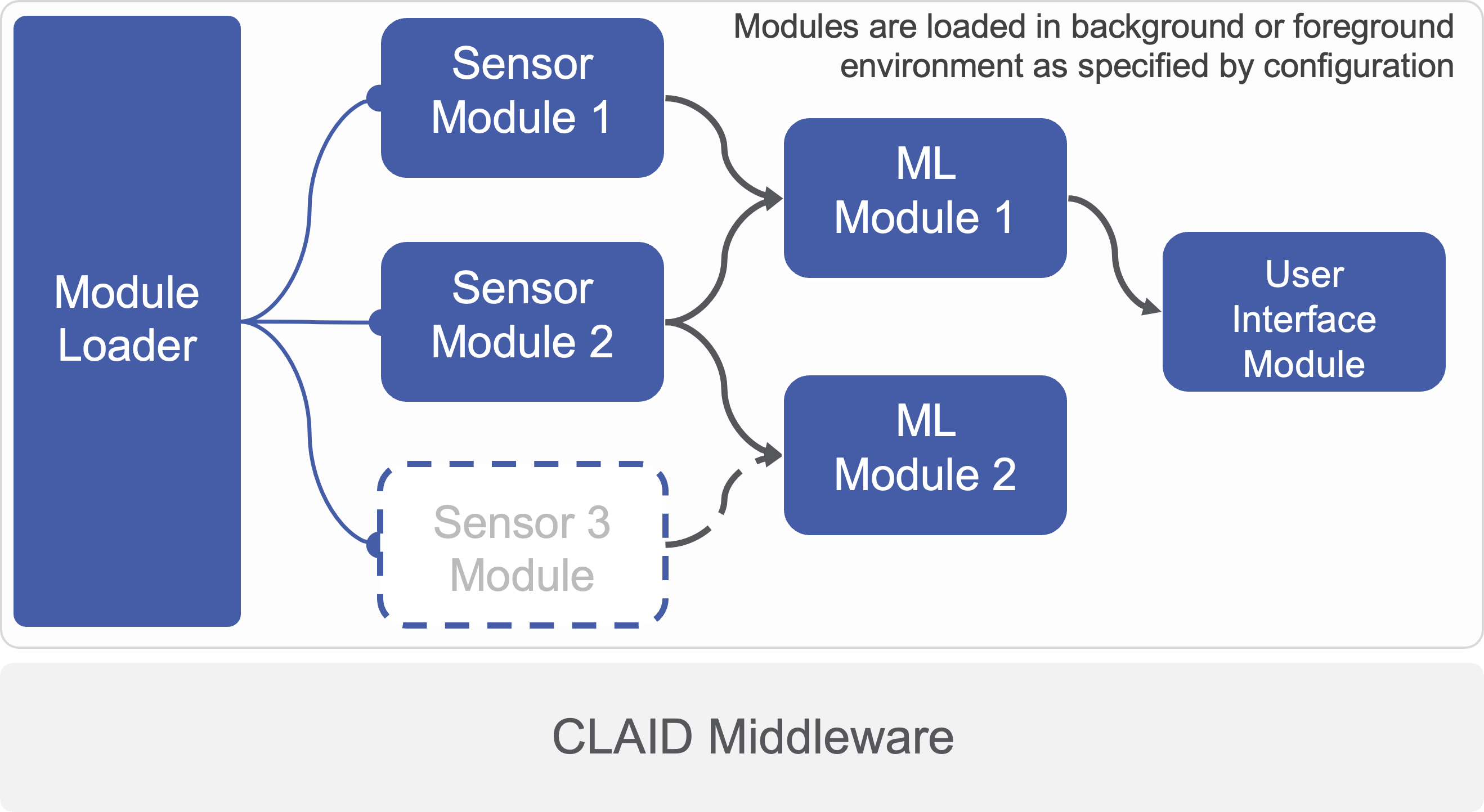}
         \caption{CLAID Modules configured for Deployment of Machine Learning Models.}
     \end{subfigure}
    \caption{Reconfiguration example, showing how we can use CLAID for data collection and model deployment. Since all implementations use loosely coupled Modules, we can reuse and connect them as required for different scenarios.}
    \label{fig:reconfiguration}
\end{figure*}
Leveraging mobile phones as mobile sensing devices is challenging. Often, developers implement data collection via monolithic pipelines which contain steps for sensor access, data serialization, and upload for each different sensor. These implementations in many cases suffer from two problems. First, these pipelines perform the individual steps synchronously one after another. Hence, it is challenging to decouple sensor data collection from data serialization and upload. If serialization takes too long or the application loses connection to the server, a delay in the processing of monolithic pipelines might lead to loss of data, if the sensor values cannot be read in time. Second, monolithic pipelines do not offer the modularity required to repurpose individual components for different scenarios, since they are strictly coupled.  After collecting data onto a server, developers can not easily change the pipeline to feed data directly into a machine learning model.

To overcome the problems of synchronous processing and the lack of modularity of monolithic pipelines, CLAID offers functionalities to compose sensor pipelines from individual and loosely coupled, asynchronous Modules. Figure~\ref{fig:reconfiguration} shows an example of how we use those Modules for scenarios of data collection and model deployment. For data serialization and upload, we provide the DataSaverModule and DataSyncModule respectively, which receive data from the sensor Modules. If all required sensor Modules are available, users of CLAID can set up data collection without programming, by providing a configuration file specifying what Modules to load and how to connect them. An example is provided by Listing~\ref{listing:claid_data_collection}, which shows how we can use the MicrophoneCollector to record audio data. This Module has properties that allow specifying recording details for the audio data and quality, such as sampling rate, encoding, bitrate, and number of channels. Furthermore, this Module provides different recording options, to configure recording start, stop, and length. In the example configuration, audio data is continuously recorded in chunks of 6 seconds, each posted separately to the "AudioData" channel. Recording starts immediately after the application has been started. Each chunk is stored in a separate WAV file by the DataSaverModule and uploaded to the Server when a connection is available.
\begin{figure}[h!t]
    \noindent\begin{minipage}[t]{\columnwidth}
    \begin{lstlisting}[caption=Example configuration for data collection,frame=tlrb,language=XML,label={listing:claid_data_collection},captionpos=b]{Name}
<Module class="MicrophoneCollector">
    <SamplingRate>16000</SamplingRate>      
    <Encoding>PCM_FLOAT</Encoding>
    <EncodingBitrate>32</EncodingBitrate>
    <Channels>MONO</Channels>
    <ContinuousChunks>6s</ContinuousChunks>
    <StartRecording>Immediately</StartRecording>
    <Output>AudioData</Output>
</Module>

<Module class="DataSaverModule">
    <save>
        <What>AudioData</What>
        <FileFormat>WAV</FileFormat>
        <StoragePath>
        <!-- files will automatically 
        be named by each Minute -->
            /sdcard/AudioData/audio_file_%M.mp3
        </StoragePath>
    </save>
</Module

<!-- Uploads data from /AudioData/ to a server, on which a DataReceiverModule is running -->
<Module class="DataSyncModule">
    <FilePath>/sdcard/AudioData</FilePath>
    <UserIdentifier>User01</UserIdentifier>
</Module>

<Module class="NetworkClientModule">
    <ConnectTo>ip:port</ConnectTo>
</Module>
    \end{lstlisting}
    \end{minipage}
\end{figure}

While it is not our goal to provide existing Modules for all possible sensors across different devices, we aim to facilitate the process of integrating new sensors via our API. As the DataSaverModule and DataSyncModule support arbitrary data types via our reflection system, integrating a new sensor only requires implementing the sensor Module using the corresponding~(native)~API of the mobile OS to read sensor values. This approach mitigates the risk of common pitfalls when implementing asynchronous sensor reading and upload. With monolithic pipelines, additional effort is required to make sure that continuous recording, for example of audio data, is ensured while uploading data in order to not block the data collection pipeline. With CLAID, those aspects are handled asynchronously via the mentioned Modules. We provide a ModuleLoader which allows running Modules in the foreground or background automatically. Furthermore, we employ different operating-system-specific techniques to ensure reliable data collection from the background via our KeepAliveModule. For example, on Android, this Module makes sure that foreground services are started in the correct way and have the necessary permissions to access sensors from the background. Further, it uses WakeLocks, to prevent the device from idling. On iOS, it makes sure that at least one background sensor is running in order to keep the application alive.
After data collection, reconfiguration allows repurposing the existing application, to now incorporate machine learning models. For example, a model for cough detection~\cite{filipe_cough_detection, filipe_cough_detection_validation} deployed on the smartphone can directly use audio data from the microphone to make predictions as shown by Listing~\ref{listing:claid_model_deployment}.
\begin{figure}
\noindent\begin{minipage}[htp]{\columnwidth}
\begin{lstlisting}[caption=Example configuration for Model Deployment,frame=tlrb,language=XML,label={listing:claid_model_deployment},captionpos=b]{Name}
<Module class="MicrophoneCollector">
    <SamplingRate>16000</SamplingRate>      
    <Encoding>PCM_FLOAT</Encoding>
    <EncodingBitrate>32</EncodingBitrate>
    <Channels>MONO</Channels>
    <ContinuousChunks>6s</ContinuousChunks>
    <StartRecording>Immediately</StartRecording>
    <Output>AudioData</Output>
</Module>

<Module class="TensorFlowLiteModule">
    <Model>CoughDetector.tflite</Model>
    <Input>AudioData</Input>
    <Output>DetectedCoughs</Output>
    <Acceleration>GPU</Acceleration>
</Module

<Module class="CoughVisualizerPlot">
    <Input>DetectedCoughs</Input>
</Module>
\end{lstlisting}
\end{minipage}
\end{figure}
In this scenario, a connection with a server is not necessarily required anymore. Our TensorFlowLiteModule allows the deployment of models directly on edge devices such as smartphones and optionally supports settings for hardware acceleration, given that a model is convertible to the TensorFlowLite format. Alternatively, our Python bindings allow us to deploy models on a connected server or other edge devices. To enable communication between Modules running on an edge device and a connected server, we leverage transparent computing. 

\subsubsection{Transparent Computing}
\label{sec:transparent_computing}
Transparent computing is a~(network) computing paradigm, that allows to logically integrate devices connected via remote connections into one system~\cite{edge_end_cloud_transparent_computing}. CLAID adopts this concept for Modules and channels. When connecting multiple instances of CLAID via a network connection, the instances exchange a list of the local and remote publishers and subscribers for each channel. If data is posted to a channel that has a remote subscriber, the data is automatically serialized to a binary format using a generic reflection system and afterward transmitted to the remotely connected instance. Upon arrival, the data is automatically deserialized and inserted into the channel. This does not differ from local communication from the perspective of the Modules, hence the involved communication is \textit{transparent} to the Modules. Modules get notified when new data is available in a channel, no matter whether this data was inserted locally or from a remote connection. Hence, regardless of the device, Modules can communicate via channels with other Modules of the logical system, without the need to use specific code for network communication. In theory, this allows arbitrary topologies of multiple devices or applications, with no hierarchy between them. However, since we use TCP/IP for network communication, some devices must act as a server, while others connect to this server as clients. By default, each CLAID instance only runs locally and does not establish any remote connection. To connect different instances, we provide the NetworkServerModule and NetworkClientModule. Those Modules allow flexible access to network interfaces. It is also possible to load multiple instances of these Modules, so that one CLAID instance can act as one or multiple Servers and/or Clients at the same time. Our implementation of transparent computing works across Android, iOS, WearOS, Linux, macOS, and Windows operating systems.

\subsubsection{Language Neutral Datatypes}
To realize transparent computing for the communication between Modules, it is necessary to serialize arbitrary data types to a binary representation to be sent via a network connection. Furthermore, this has to work across different programming languages, so that a Module written in Python, for example, can receive data from a Module written in Java and vice versa. To achieve this, we require platform-agnostic and cross-language data types.
These can typically be created via Interface Description Languages~(IDLs) such as protobuf~\cite{protobuf} or the IDL used by ROS~\cite{ros}. However, IDLs require code generation and do not allow defining functions for data types. They mostly are containers to store and pass data. With this approach, it is not easily possible to use existing data types of applications for transparent computing, or only to an extent limited to primitive data types.
To overcome these limitations, we developed a custom reflection system, inspired by those used in some robotic middleware frameworks such as MIRA~\cite{mira}. This reflection system allows the integration, serialization, and deserialization of arbitrary data types, as well as enabling function calls in all supported languages. Furthermore, our implementation of a reflection system enables us to automatically generate data type bindings for different programming languages, for example using pybind11~\cite{pybind11} and jbind11~\cite{jbind11}, respectively.

\subsubsection{Offloading Capabilities}
Current mobile and edge devices often do not have enough resources in terms of computing and memory capacities to execute machine learning models and algorithms for analysis. Using our transparent computing approach (cf. section~\ref{sec:transparent_computing}), CLAID enables static offloading among multiple connected devices on the Module level. Modules that need a task to be performed by a more powerful device can post data to a corresponding channel. If a Module running on a remotely connected device can process data from this channel, it runs the necessary calculations and posts the results to a different channel. The communication between the Modules is handled transparently. Since our implementation is compatible with all major operating systems, offloading works across different devices. Currently, we support static offloading~\cite{static_and_dynamic_offloading} which requires manual configuration via configuration files. We highlight some use cases in the following.

\paragraph{Smartphone to Server}
While we provide Modules to directly deploy machine learning models on mobile devices, for example via our TensorFlowLiteModule, some models can not be run using the same approach due to the limited computation power of edge devices. In such cases, CLAID supports hosting the models on a connected Server, either via the TensorFlowLiteModule or by using Python bindings to integrate existing machine learning pipelines. Machine learning Modules can subscribe to the data channels of the sensor Modules. Using this approach, it is possible to stream data from the smartphone directly to a Python application running on a connected device. This allows the implementation and deployment of distributed data analysis pipelines across devices, enabling the use of various Python packages for data processing and machine learning.

\paragraph{Smartwatch to Smartphone}
Smartwatches are able to collect data of different modalities, such as heart rate, oxygen saturation, physical activity, body composition, and more \cite{werable_technology_survey, wearables_mental_health_survey}. Machine learning-based approaches can derive novel medical insights from such smartwatch data, possibly enabling continuous tracking of a person's health state. For example, physical activity data recorded by smartwatches can potentially allow the development of novel digital biomarkers for inflammation and mortality~\cite {jinjoo_nature}. Since smartwatches have limited processing capabilities, they can not run the required analysis, such as inference of a machine learning model, directly. To overcome this limitation, CLAID supports WearOS and enables offloading of tasks to nearby devices such as smartphones, connected via Bluetooth. Additionally, our framework also allows offloading from smartwatches to a server, if Wi-Fi or cellular connectivity is available. Next to Android Modules, which allow data collection on WearOS as well, we provide additional sensor Modules for smartwatches like the Samsung Galaxy Watch~\cite{samsung_galaxy_watch}.

\subsection{Experiments}
\label{sec:experiments}
In the following, we describe the experiments we conducted in order to evaluate CLAID. 
With our first experiment, we assess the system's ability to collect data reliably on smartphones, employing both internal and external sensors. Internal sensors refer to internal smartphone sensors, like GPS or microphones, while external sensors refer to wearables that are paired via Bluetooth, like a Smartwatch. For the second experiment, we propose a novel verification method for machine learning models deployed on edge devices, which we term \textit{ML-Model in the Loop}. For this, we deploy a model for cough detection~\cite{filipe_cough_detection, filipe_cough_detetion_validation_2023} on smartphones and evaluate it by sending test data from the server. Lastly, we assess memory and battery consumption. All experiments are performed on Android and iOS respectively, using the devices shown in table~\ref{tab:used_devices}.
\begin{table}[h!t]
\caption{Overview of the devices we used for conducting the experiments.}
\label{tab:used_devices}
\centering
\begin{tabularx}{\columnwidth}{p{1.6cm}YYYp{2cm}}
Device          & OS & RAM  & Battery & CPU                     \\ \hline
Xiaomi Redmi 10 & Android 11       & 4GB    & 5000 mAh & Qualcomm Snapdragon 680 \\ 
\hline
iPhone 11       & iOS 16.1.2         & 4GB           & 3110 mAh & Apple A13 Bionic        \\ \hline
\end{tabularx}
\caption*{This table shows the devices we used for the experiments as well as their corresponding specifications, such as random access memory~(RAM) capacity, central processing unit~(CPU), and internal storage.}
\end{table}
\subsubsection{Sampling Coverage}
\label{sec:experiments:sampling_coverage}
Consistent with prior work~\cite{carp_cams}, we conducted an experiment to test CLAID's data collection capabilities. In this experiment, we collected data over 24 hours from multiple sensors with fixed sampling rates. From the sampling rate, we can calculate the number of measurements that are expected each hour in order to create a metric for the percentage of data collected per hour. For example, accelerometer data sampled at 20Hz, results in $20 \cdot 60 \cdot 60 = 72000$ samples per hour. We focus on passive data collection for this experiment, using sensors that run continuously without further interaction. We use internal and external sensors with configured sampling frequencies as listed in Table~\ref{tab:sampling_coverage_sensors}.
\begin{table}[h!t]
\renewcommand{\arraystretch}{1.2}
\caption{Overview of the used sampling frequencies for each sensor for the sampling coverage experiment. }
\label{tab:sampling_coverage_sensors}
\begin{tabularx}{1.0\columnwidth}{YYYYY}
\hline
 \multicolumn{5}{c}{Smartphone sensors~(Internal)}\\\hline
Accel. & Battery level & Connec- tivity & Location  & Microphone audio  \\\
 20Hz & $\frac{1}{60}$Hz & 1Hz & $\frac{1}{20}$Hz & $\frac{1}{10}$Hz \\ \hline \hline
 \multicolumn{3}{c}{Polar Verity Sense~(External)} & \multicolumn{2}{|c}{CORE~(External)}\\\hline 
 Accel. & Battery level & Heart rate & \multicolumn{2}{|c}{Core body temperature}\\
 20Hz & $\frac{1}{60}$Hz & $\frac{1}{10}$Hz & \multicolumn{2}{|c}{ 1Hz}\\\hline
 \end{tabularx}
 \caption*{We use~5 internal sensors that are integrated with the smartphone, and two external sensors, namely the Polar Verity Sense and the CORE body temperature sensor. Data from each sensor is requested at defined interval rates as shown in the table.}
 \renewcommand{\arraystretch}{1.0}
\end{table}
We chose five internal smartphone sensors~(i.e., accelerometer, connectivity, location, microphone, and battery level) and two external sensors, namely a Polar Verity Sense~\cite{polar_verity} and a CORE body temperature sensor~\cite{core_sensor}.  Audio data from the microphone is recorded at 16kHz, in continuous chunks of 10 seconds. In other words, one audio file was recorded every 10 seconds, with each file being 10 seconds in length, resulting in continuous recording over 24 hours. For this, we verify a continuous recording of the audio data by appending each of the 8640 recorded files in order to produce one coherent audio file for 24 hours by screening it for discontinuities in the recordings. The other internal sensors were configured to provide data at the sampling rates as specified in the table.

By using external sensors, we aim to demonstrate how it is possible to integrate new and custom sensors with CLAID, here using the Polar Verity Sense and the CORE sensor. The Polar Verity Sense is an optical heart rate monitor in the form of a wrist strap and features, among others, heart rate monitoring and accelerometer recording capabilities. The device transmits data to a smartphone via a Bluetooth connection. It comes with an SDK, aiding integration into applications for Android and iOS. The CORE sensor is a non-invasive core body temperature sensor. It allows measuring core body temperature by attaching the sensor to a person's body~\cite{core_validation_study}. The CORE sensor, too, can transmit data via Bluetooth, however, no public SDK is provided. Integration of the sensor has to be implemented manually. With the choice of this sensor, we want to demonstrate that how CLAID facilitates the integration of unknown sensors that do not have a specific SDK. The external sensors were worn by one of the authors of this paper over the course of the experiment. The Polar Verity Sense was worn on the upper arm using the included belt. The CORE device was attached to the body on the rib cage over the latissimus muscle, as instructed by the user manual.
In addition to investigating sampling coverage as a metric of how much data was collected per hour, we want to evaluate the reliability of data upload during fluctuating network connectivity. We controlled the smartphone's network connection using a predefined protocol over a 24-hour period and examined how CLAID responded to long disconnects and switching from Wi-Fi to Cellular or vice versa. Table~\ref{tab:sampling_coverage_connectivity_protocol} shows the specified protocol.
\begin{table}[h!t]
\centering
\caption{Connectivity protocol for the sampling coverage experiment.}
\label{tab:sampling_coverage_connectivity_protocol}
\begin{tabularx}{0.35\textwidth}{cc}
\hline
Hours             & Connectivity mode  \\\hline
00:00 - 02:00    & Cellular            \\
02:00 - 10:00    & Wi-Fi                \\
10:00 - 12:00    & Disconnected        \\
12:00 - 18:00    & Cellular            \\
18:00 - 20:00    & Disconnected        \\
20:00 - 22:00    & Wi-Fi               \\
22:00 - 00:00    & Cellular            \\\hline                     
\end{tabularx}
\caption*{This table shows the connectivity protocol we followed for the sampling coverage experiment. The connectivity mode refers to the network modality currently available for the smartphone and can either be a Wi-Fi connection, cellular, or disconnected.}
\end{table}
This protocol enables us to examine transitions between Wi-Fi and Cellular networks. When a connection is available, CLAID transmits recorded data to the server. Otherwise, the system stores recorded data on the phone's internal memory until a connection is re-established, and synchronizes it afterward. We study data synchronization after prolonged disconnects, once through Wi-Fi and once through Cellular connections. 

\subsubsection{ML-Model in the Loop: Offloading for the verification of deployed Machine Learning Models}
\label{sec:exeriments:ml_evaluation}
Verifying and validating machine learning models in different scenarios, for example for medical use cases, requires deploying them in a way that they can use data recorded via sensors on the edge device in order to enable continuous analysis. Ideally, these models could be deployed directly on edge devices like smartphones. When transferring a model from the controlled training and test environment on a server to the edge, however, some problems might arise. To run the models on the edge device, it might be necessary to convert them to a supported format (e.g., TensorFlowLite) or to apply quantization, which can lead to differences in the model performance. Furthermore, differences in internal sensors between different devices can impact the accuracy and behavior of machine learning models. Differences in hardware and operating systems of devices can also lead to variations in calculations and result in varying model outputs\cite{floating_point_calculations_across_platforms}. Due to these occurring variances, one has to expect that a model performs differently when first deployed in the field, compared to the lab environment. To investigate variances in a model across devices and to verify the correct execution of the model, we propose a novel methodology termed \textit{ML-Model in the Loop}, that allows migrating models from the lab environment step-by-step and in a controlled manner, which is inspired by software-in-the-loop approaches often used for analytical verification~\cite{software_in_the_loop}. When first deploying the model on the phone, we can test its performance by sending known-good test data from a server and comparing the results with the ground truth. This allows for investigating whether the implementation of the model on the smartphone is correct for data where a ground truth is available, and to investigate differences step-by-step. 
\begin{figure}[h!t]
  \centering
  \includegraphics[width=1.0\linewidth]{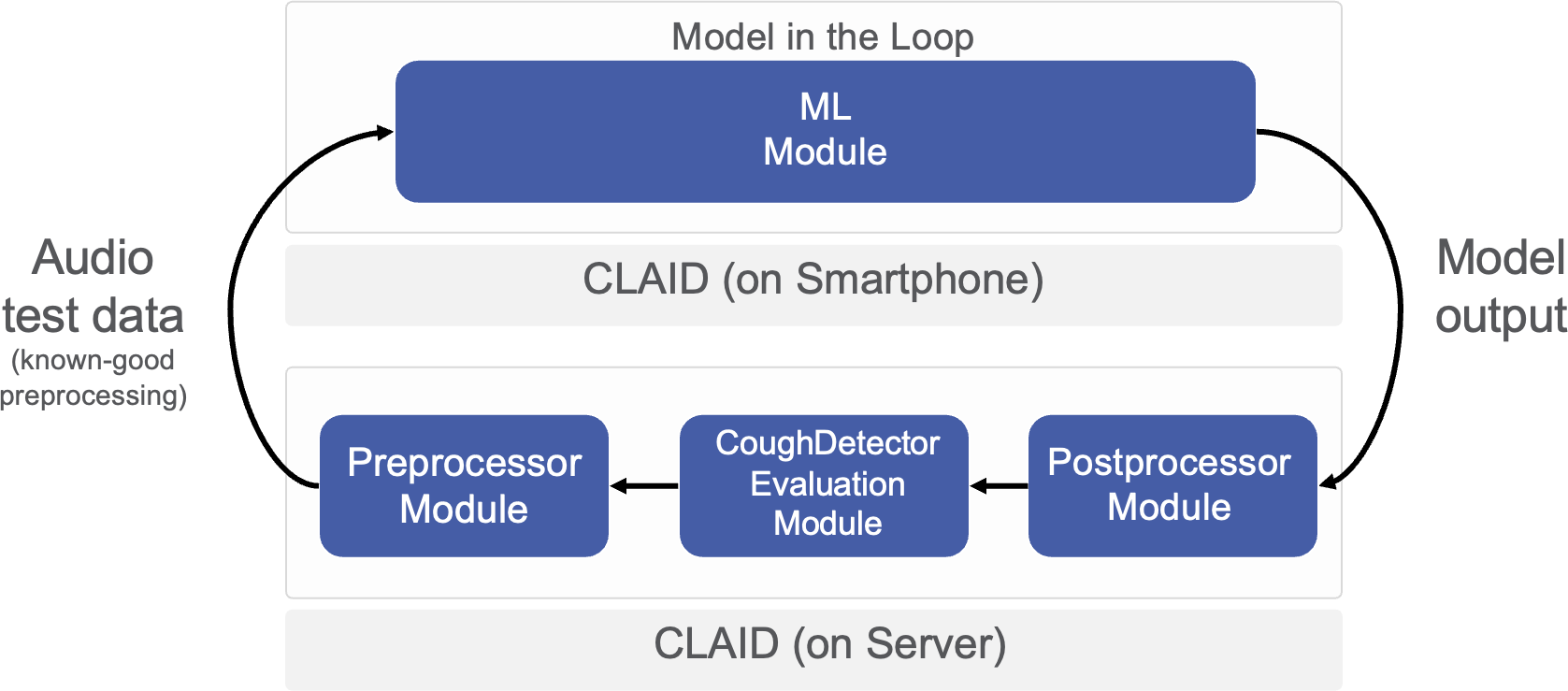}
  \caption{This figure shows our proposed setup for the verification of ML models on edge devices, which we refer to as \textit{ML-Model in the Loop}. In This scenario, known-good test data from the server is streamed to the model on a connected device and evaluated to ensure correct implementation.}
  \label{fig:experiments:ml_deployment:model_under_test}
\end{figure}
This novel method for a verification experiment is enabled by CLAID's transparent computing mechanism. In many cases, training of machine learning models is done with Python using one of the major machine learning frameworks (e.g., TensorFlow). An implemented pipeline for the inference of a model typically consists of code for preprocessing, model inference, and postprocessing~\cite{data_science_pipelines}. Python is supported by CLAID. Therefore, Python Modules running on a Server can communicate with Modules running on a smartphone. This integration allows verification of individual parts of the pipeline, separately and isolated when transferring the model from a training environment on a server to an edge device. In our proposed scenario, the model would be trained on the server and afterward deployed on the smartphone, as seen in Figure~\ref{fig:experiments:ml_deployment:model_under_test}. To make sure the model performs correctly on the smartphone, especially after conversion and quantization, we can test it using known-good and preprocessed data from the server using existing code in Python. Afterward, pre- and postprocessor Modules can be adapted for deployment on the Smartphone as well and tested one after another. As a final step, all Modules can run on the edge device and receive known-good test data from the server. If the model performs well compared to the original implementation on the server, it can directly work with data received from the smartphone sensors. The remaining variance in performance can then be attributed to differences in hardware sensors.
\begin{figure}
\noindent\begin{minipage}[t]{\columnwidth}
\begin{lstlisting}[caption=CoughDetectorEvaluator Module in Python example,frame=tlrb,language=Python,label={listing:results:claid_python_cough_detector_evaluator_module},captionpos=b]{Name}
from PyCLAID import PythonModule as Module
from PyCLAID import PyLayerData as LayerData

class CoughDetectorEvaluator(Module):
    def initialize(self):
        self.audio_channel 
            = self.publish(AudioData, "AudioData")
            
        self.result_channel_a = 
            self.subscribe(LayerData, 
                model_a_channel_name, onData)

        audio_sample = AudioData("cough_01_test.mp3")
        
        # Audio data posted by the PythonModule on 
        # the Server will be forwarded to the Smartphone.
        self.audio_channel.post(audio_sample)

    # Called, when output of the ML model running on
    # the smartphone becomes available.
    def onData(self, data):
        model_output = np.array(data, copy=False)
        # ... compare model output with ground truth
        # ... post next audio sample
\end{lstlisting}
\end{minipage}
\end{figure}
In our experiment, we demonstrate this method using the cough detection ensemble by Barata et. al~\cite{filipe_cough_detection, filipe_cough_detection_validation}. This algorithm analyzes 6-second audio segments by applying a sliding window approach, which calculates a Mel spectrogram for each window. The number of windows per file can differ since some windows might be ignored due to silence. The windows are then processed by an ensemble of five cough detection models to predict cough probabilities, from which we calculate the number of coughs over 6 seconds. For further details, we refer to the referenced publications. In the experiment, we run the ensemble on both Android and iOS and send test data from the server by integrating existing evaluation code as a Python Module, which can communicate with the model on the phone via channels. We compare the results regarding the two platforms in terms of the number of coughs detected and window probabilities calculated. In this experiment, we are interested in comparing the model implementations between Android and iOS using an arbitrary test set. We do not evaluate the model performance as a cough detector in general, as this has been done in previous work~\cite{filipe_cough_detection_validation}. The test data consists of 278 audio files with a length of 6 seconds each, recorded from 6 different patients~(5 male, 1 female). In total, 281 coughs occurred, with some files containing 0 and others up to 6 coughs. 

\subsubsection{Memory and battery consumption}
\label{sec:experiments:memory_battery}
To investigate if CLAID's flexibility and abstractions come at a cost of increased resource usage, we conducted an experiment that allows us to determine resource utilization in terms of battery and memory consumption, as well as to understand the impact on the overall performance of the device it is running on. This information is useful for making decisions about the suitability of CLAID for a particular application or usage scenario. Comparing memory and battery consumption between Android and iOS devices can provide valuable insights into how the performance varies across different operating systems and hardware configurations. For the experiment, we use a subset of the sensors used in the sampling coverage experiment presented in section~\ref{sec:experiments:sampling_coverage}, namely the accelerometer, location, and connectivity. The accelerometer and location sensors are both hardware-based and are expected to consume more resources compared to the software-based connectivity sensor. They were chosen over other hardware sensors because they are two of the most commonly used and resource-intensive sensors of smartphones. We first run each sensor individually for two hours and afterward in combination. We then compare the results to the baseline, where CLAID runs without any sensors enabled, to gain insights into the scalability of CLAID when using multiple sensors. Within the two-hour time frame, we expect the battery consumption per sensor to be approximately linear. It is therefore possible to make an approximate estimation of the current draw for each sensor by calculating the slope of the graph that represents the battery consumption during this period. 
This allows estimating the \textit{expense} of running individual sensors in terms of current draw.
\section{Results}
\label{sec:results}
We present results as well as important findings of the experiments described in Section~\ref{sec:experiments} in the following. We discuss the sampling coverage experiment in Section~\ref{sec:results:sampling_coverage} and the ML-Model in the Loop Experiment in Section~\ref{sec:results:ml_deployment}, respectively. We discuss findings regarding memory and battery consumption in Section~\ref{sec:results:battery_memory}.
\subsection{Sampling coverage}
\label{sec:results:sampling_coverage}
\begin{figure}[h!t]
     \centering
     \begin{subfigure}{1.0\columnwidth}
         \centering
         \includegraphics[width=1.0\columnwidth]{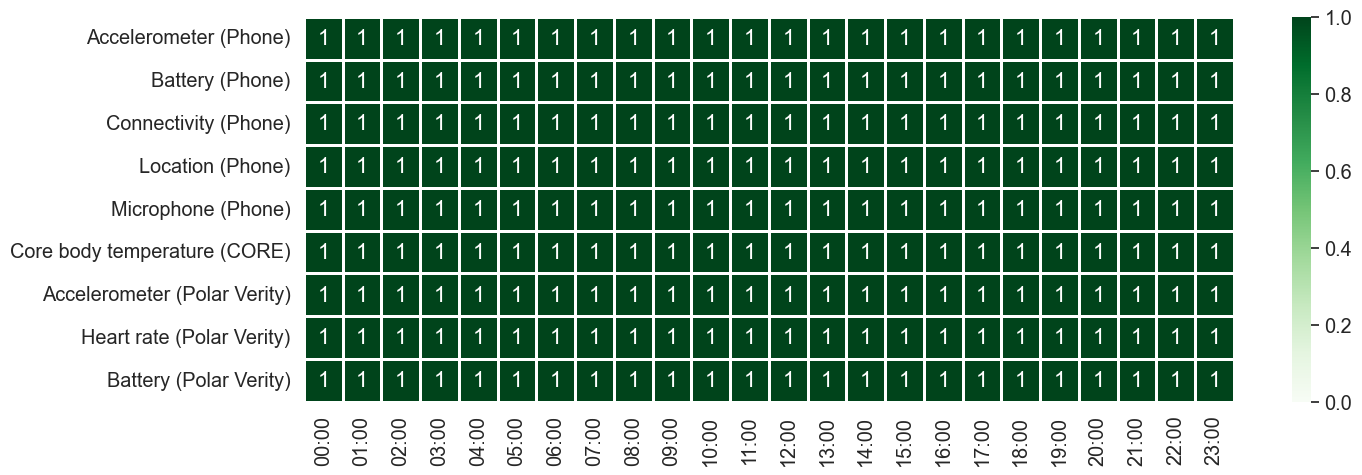}
         \caption{Sampling coverage over 24 hours~(Android)}
     \end{subfigure}
     \begin{subfigure}{1.0\columnwidth}
         \centering
         \includegraphics[width=1.0\columnwidth]{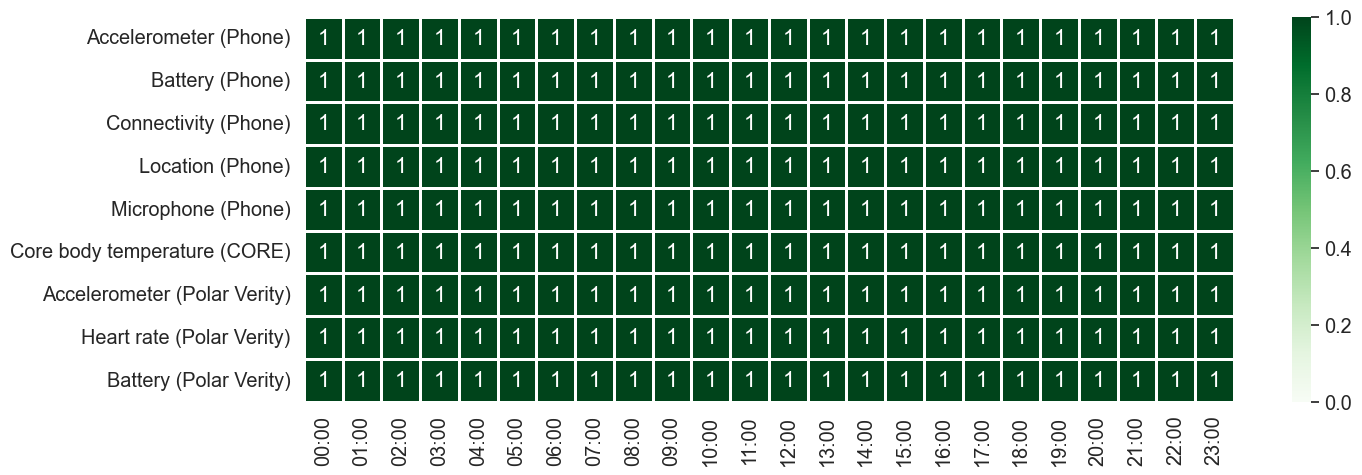}
         \caption{Sampling coverage over 24 hours~(iOS)}
     \end{subfigure}
    \caption{Outcome of the sampling coverage experiment. Data is gathered from various sensors for 24 hours on both Android and iOS using the exact same sampling configuration. The sampling rate determines the predicted number of samples per hour.}
    \label{fig:results:sampling_coverage}
\end{figure}

\begin{figure}[h!t]
     \begin{subfigure}[b]{0.32\columnwidth}
         \centering
         \includegraphics[width=1.0\columnwidth]{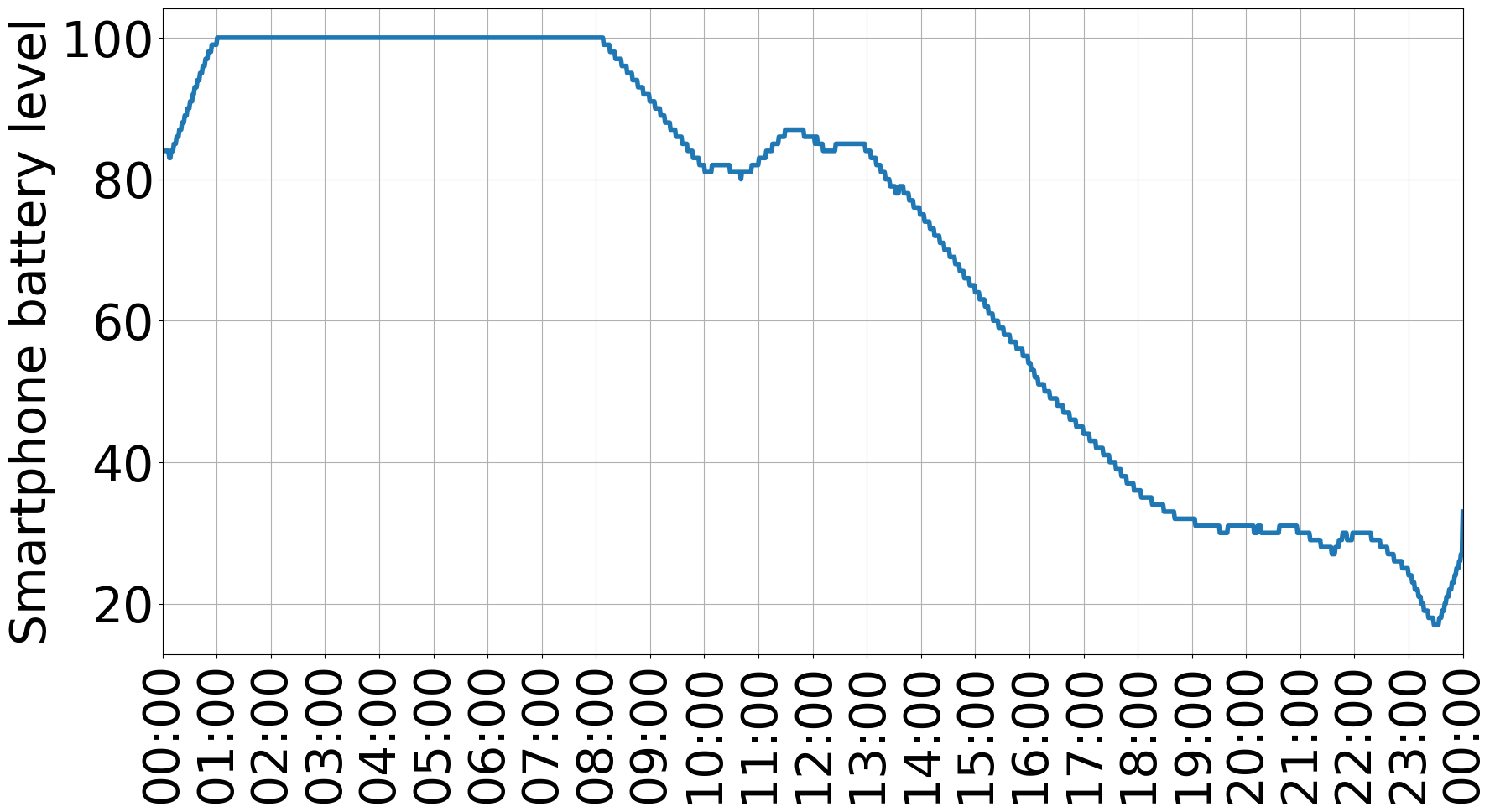}
         \caption{Smartphone battery level}
         \label{fig:results:smartphone_battery}
     \end{subfigure}
     \begin{subfigure}[b]{0.32\columnwidth}
         \centering
         \includegraphics[width=1.0\columnwidth]{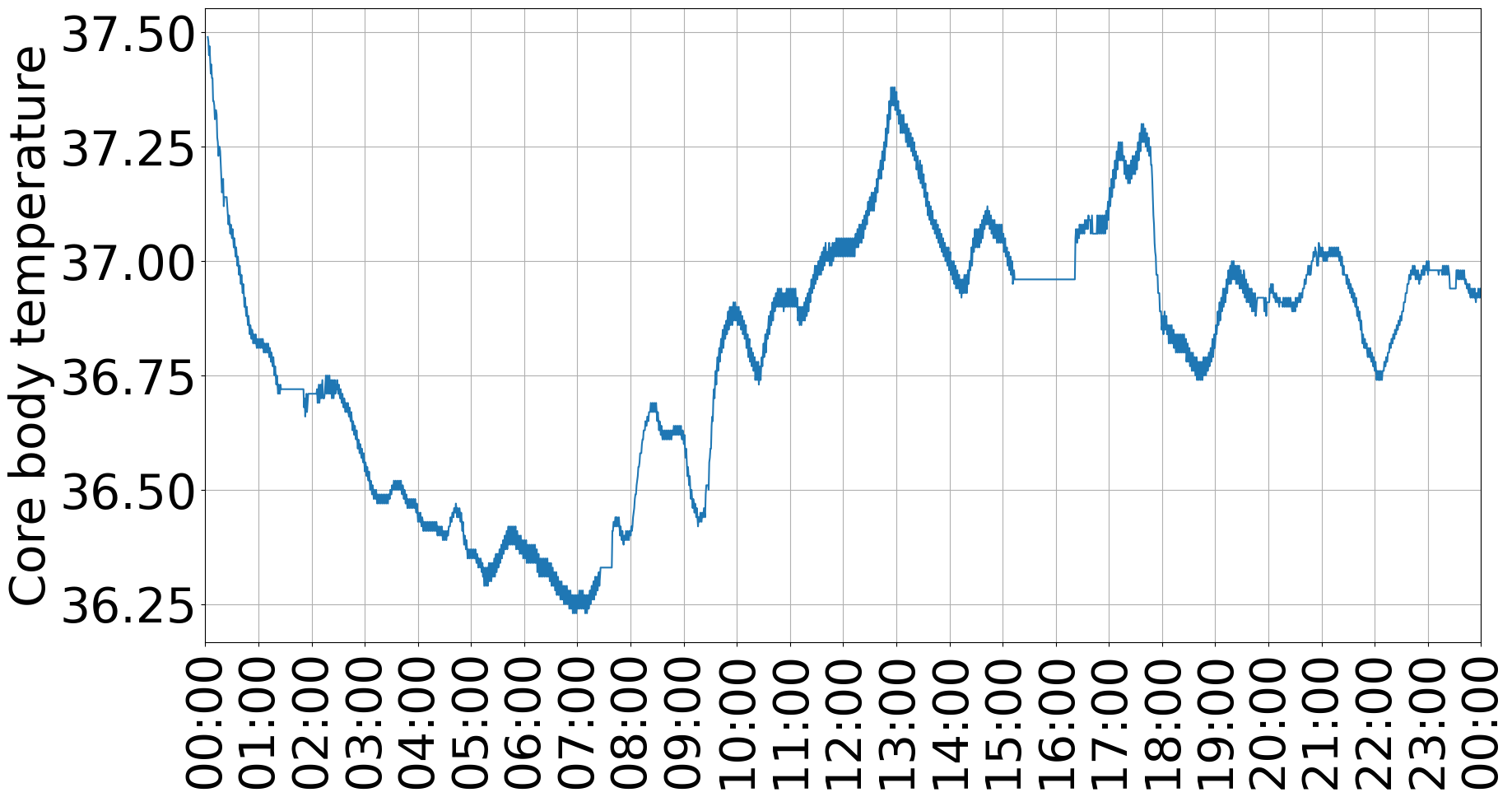}
         \caption{Core body temperature}
         \label{fig:results:core_body_temperature}
     \end{subfigure}
     \begin{subfigure}[b]{0.32\columnwidth}
         \centering
         \includegraphics[width=1.0\columnwidth]{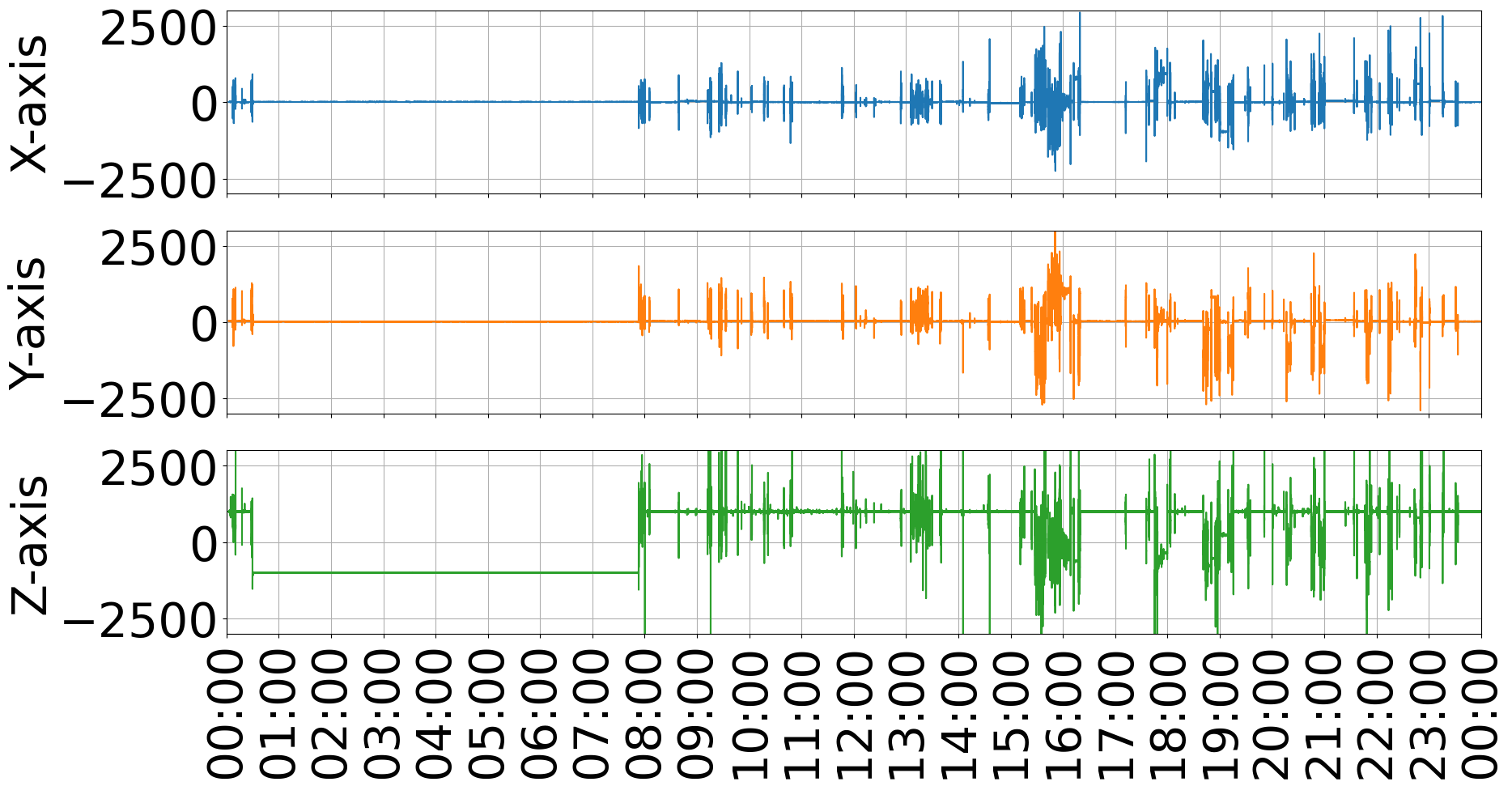}
         \caption{Smartphone accel. 20Hz}
         \label{fig:results:smartphone_accelerometer}
     \end{subfigure}
     \begin{subfigure}[b]{0.32\columnwidth}
         \centering
         \includegraphics[width=1.0\columnwidth]{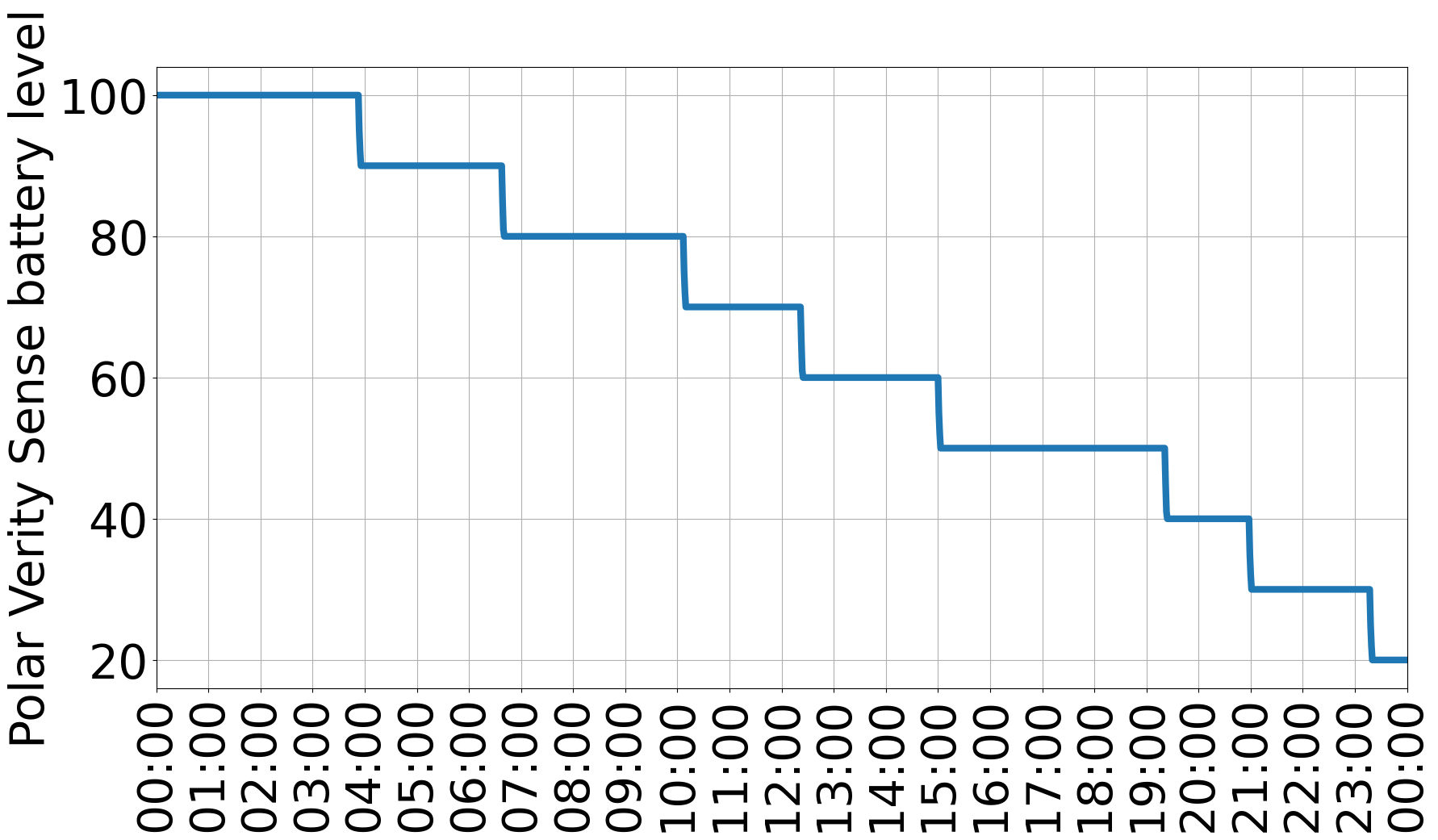}
         \caption{Polar Verity battery level}
         \label{fig:results:polar_battery}
     \end{subfigure}
     \begin{subfigure}[b]{0.32\columnwidth}
         \centering
         \includegraphics[width=1.0\columnwidth]{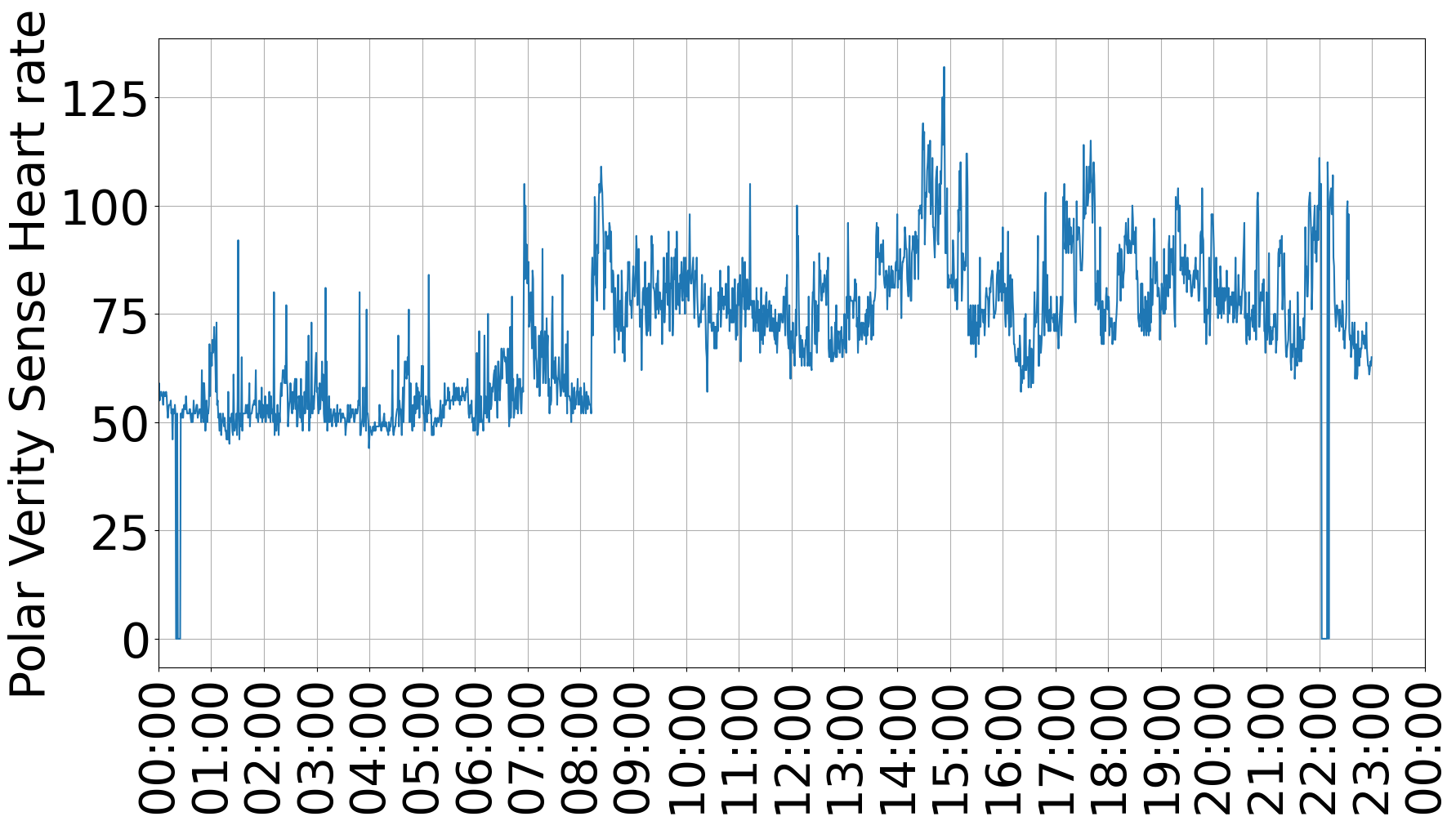}
         \caption{Polar Verity heart rate}
         \label{fig:results:heart_rate}
     \end{subfigure}
     \begin{subfigure}[b]{0.32\columnwidth}
         \centering
         \includegraphics[width=1.0\columnwidth]{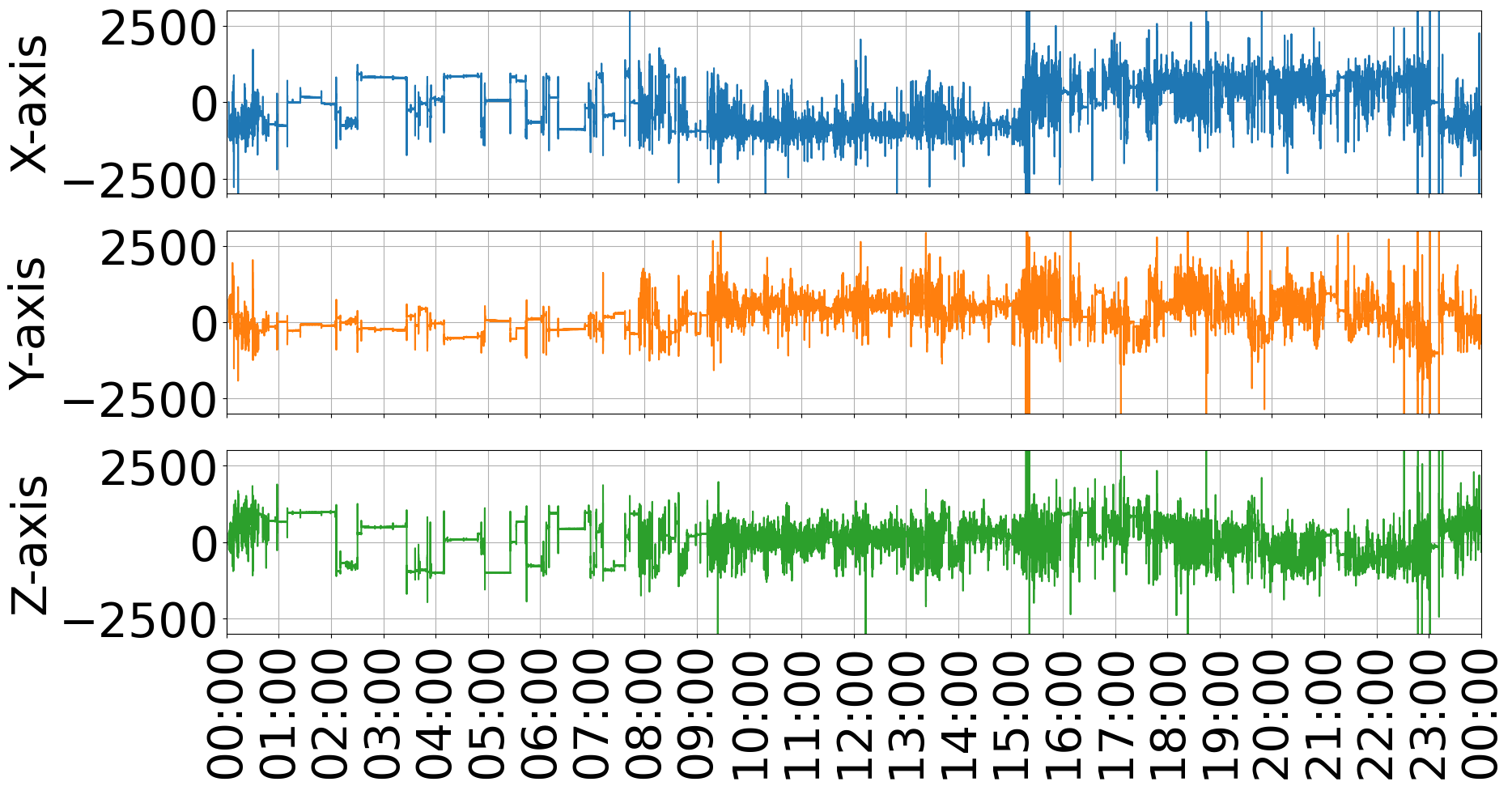}
         \caption{Polar Verity accel. 20Hz}
         \label{fig:results:polar_accelerometer}
     \end{subfigure}
        \caption{Example plots of data collected over 24 hours. }
    \label{fig:results:example_plots}
\end{figure}

\begin{figure*}[h!t]
     \begin{subfigure}[b]{0.49\textwidth}
         \centering
         \includegraphics[width=1.0\textwidth]{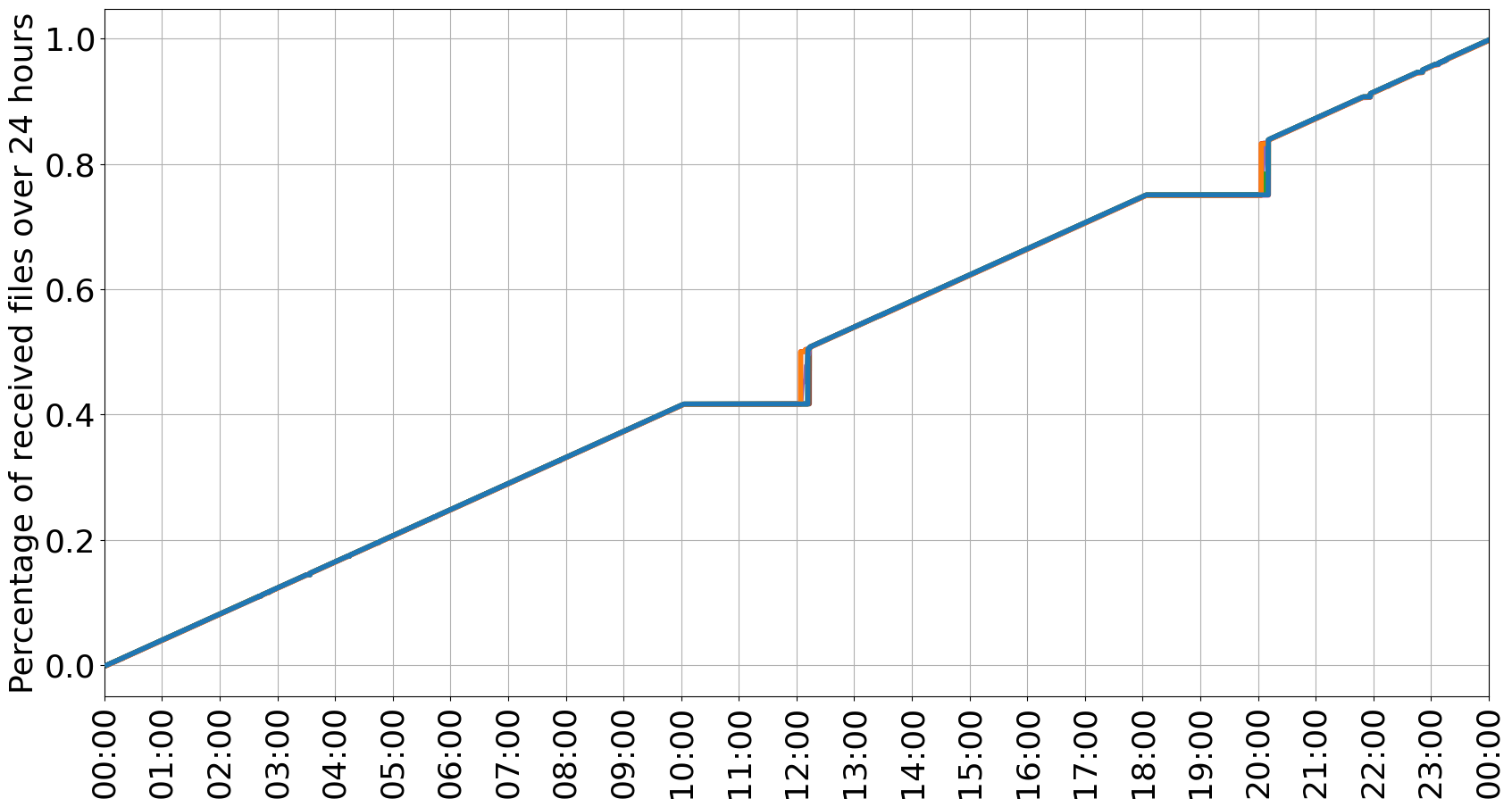}
         \caption{Data arrival times~(Android)}
         \label{fig:results:sampling_coverage_arrival_times:android}
     \end{subfigure}
     \begin{subfigure}[b]{0.49\textwidth}
         \centering
         \includegraphics[width=1.0\textwidth]{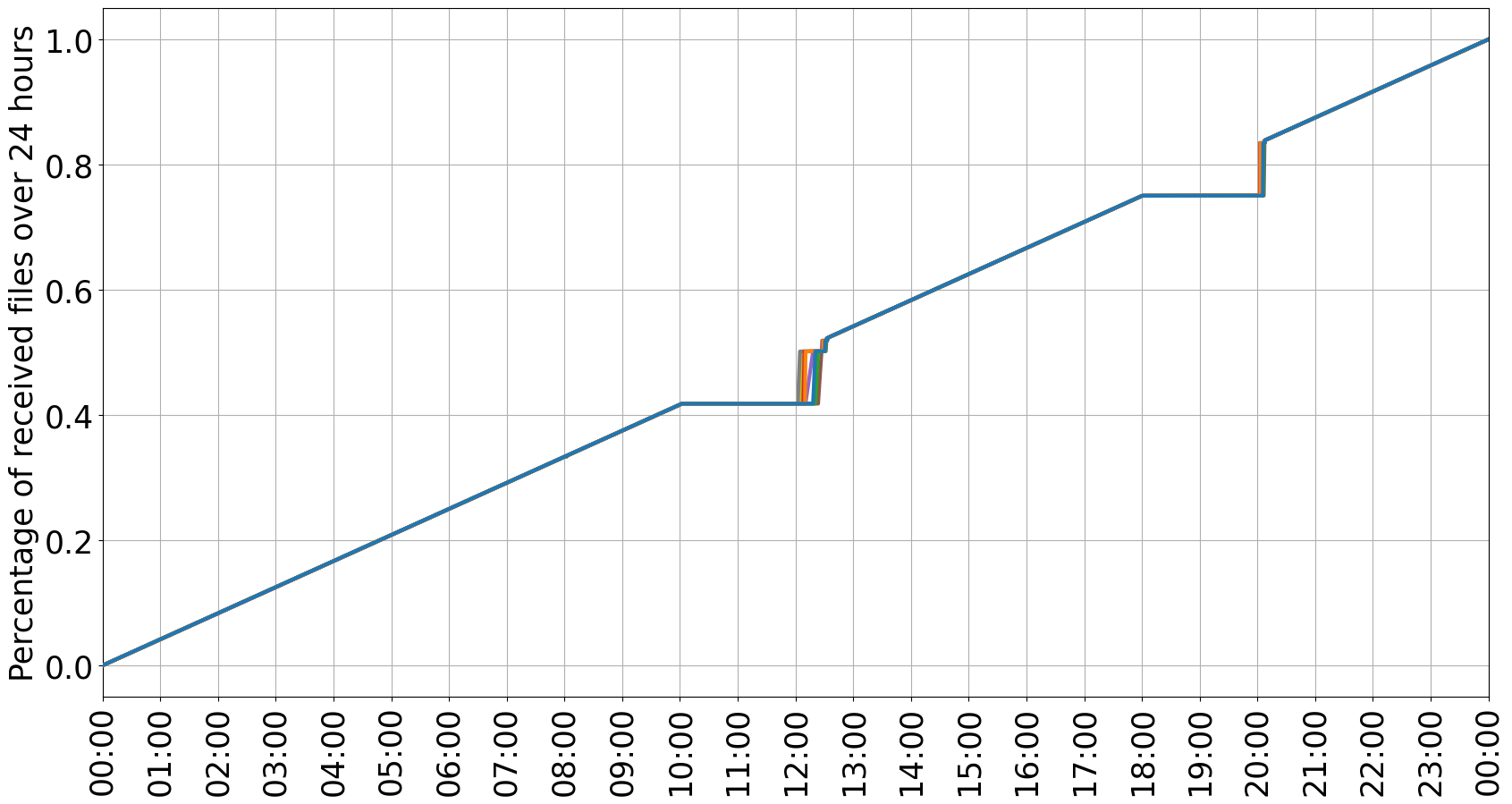}
         \caption{Data arrival times~(iOS)}
         \label{fig:results:sampling_coverage_arrival_times:ios}
     \end{subfigure}     
     \begin{subfigure}{0.49\textwidth}
         \centering
         \includegraphics[width=1.0\textwidth]{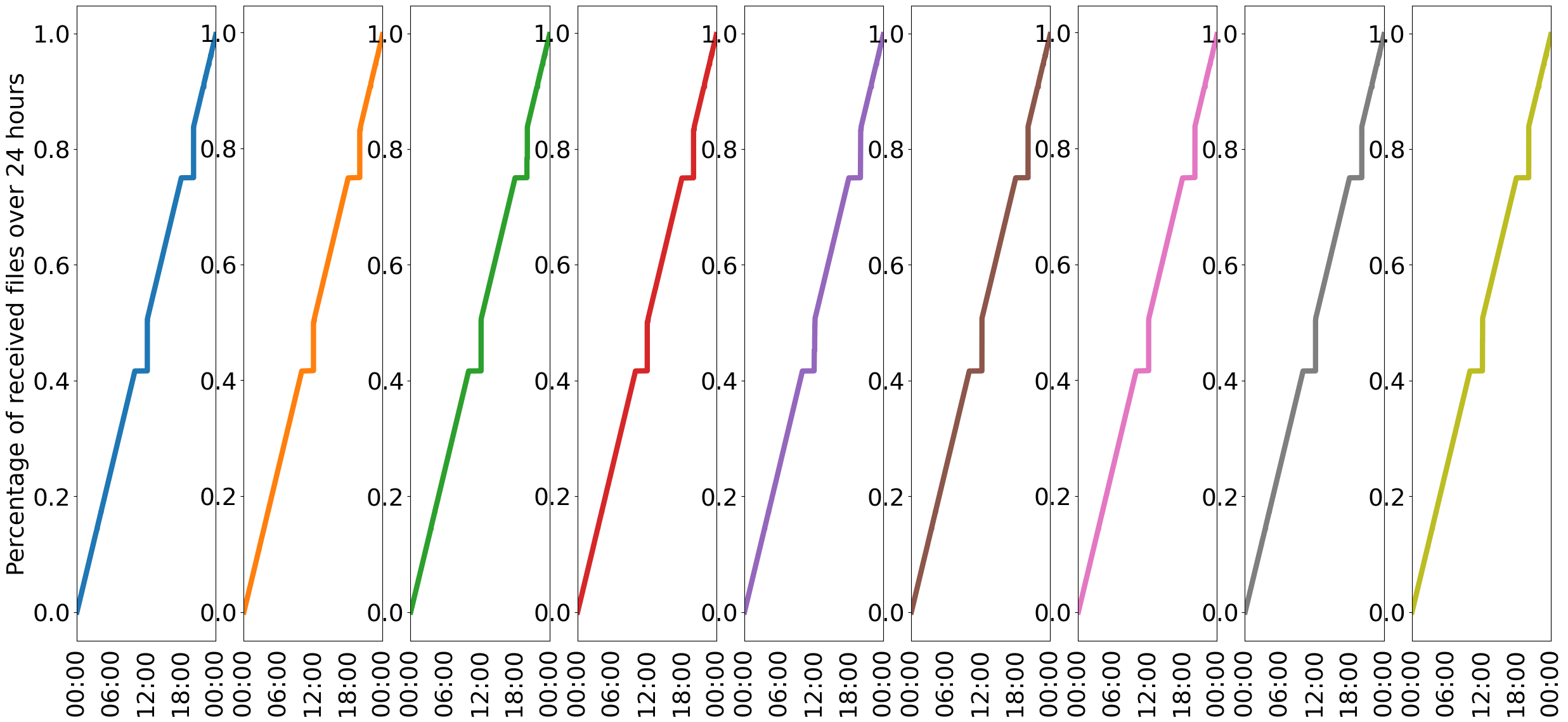}
         \caption{Data arrival times separate per sensor~(Android)}
         \label{fig:results:sampling_coverage_arrival_times:ios_parallel}
     \end{subfigure}
 \begin{subfigure}{0.49\textwidth}
         \centering
         \includegraphics[width=1.0\textwidth]{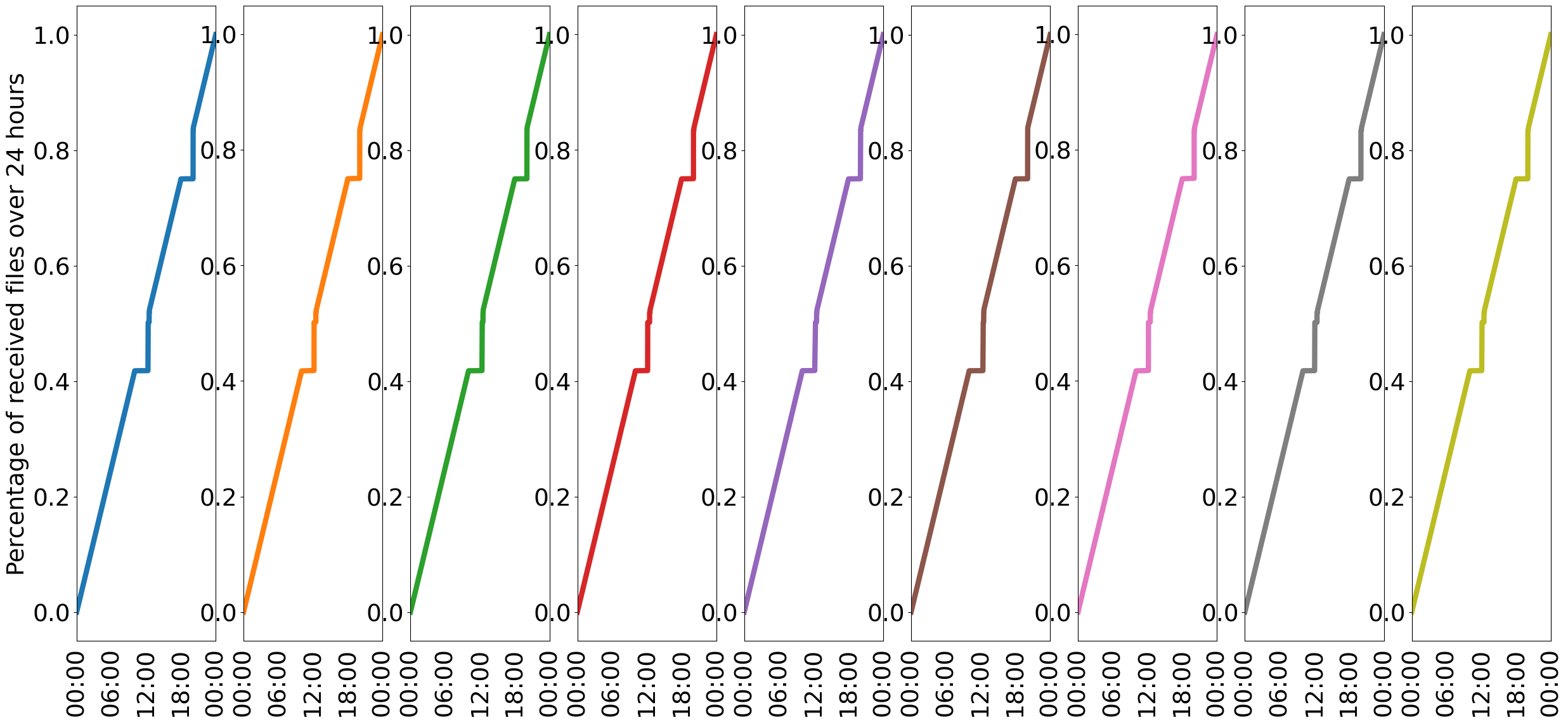}
         \caption{Data arrival times separate per sensor~(iOS)}
         \label{fig:results:sampling_coverage_arrival_times:android_parallel}
     \end{subfigure}
      \begin{subfigure}{1.0\textwidth}
         \centering
         \includegraphics[width=1.0\textwidth]{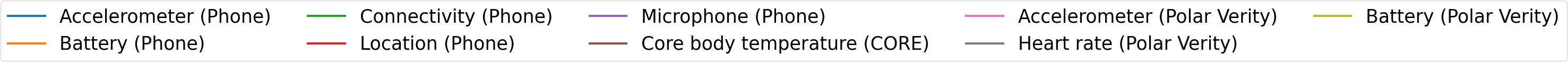}
     \end{subfigure}
        \caption{Data arrival times of samples for each sensor. We normalize the percentage of already received samples for each sensor based on the total number of samples expected after 24 hours. Discontinuities represent disabled network connection between 10:00-12:00 and 18:00-20:00}
        \label{fig:results:sampling_coverage_arrival_times}
\end{figure*}

Figure~\ref{fig:results:sampling_coverage} shows the results of the sampling coverage experiment. We visualize the percentage of samples received for each sensor and for each hour. 
As can be seen, we achieved 100\% sampling coverage for each sensor, namely the accelerometer, battery level, location, connectivity, and microphone data on the smartphone, the accelerometer, heart rate, and battery level on the Polar Verity, as well as core body temperature from the CORE device, on both Android and iOS. We visualize some examples of data recorded over the 24 hours on the Android device in Figure~\ref{fig:results:example_plots}.
The battery level of the smartphone is shown in Figure~\ref{fig:results:smartphone_battery}. The smartphone was plugged in for charging over the night from 00:30 a.m. to 08:00 a.m., leading to a full charge of the battery. It was used during the day with additional charging around 11:00 a.m. and 9:30 p.m. The battery was drained continuously between 11:00 a.m. and 9:30 p.m., as shown by the Figure. Analogously, the battery level of the Polar Verity Sense is shown in Figure~\ref{fig:results:polar_battery}. The device was fully charged when we started the experiment at 00:00 a.m. and continuously used for 24 hours, leading to a decrease in the battery level from 100\% to 20\% over the course of 24 hours. The measurements of the core body temperature are shown in Figure~\ref{fig:results:core_body_temperature}. We measured a maximum temperature of 37.49 degrees Celsius at 00:00 a.m., directly after putting the sensor into operation, and a minimum temperature of 36.23 degrees at 07:00 a.m. The discontinuities between 02:00 p.m. and 03:00 p.m. result from temporarily removing the sensor from the body. The heart rate measured by the Polar Verity Sense is shown in Figure~\ref{fig:results:heart_rate}. Discontinuities around 00:20 a.m. and 10:20 p.m. result, again, from temporarily removing the sensor from the body. We measured a minimum heart rate of 44.0 beats per minute~(bpm) at around 04:00 a.m. and a maximum heart rate of 133~bpm at around 03:00 p.m. Lastly, Figures~\ref{fig:results:smartphone_accelerometer} and~\ref{fig:results:polar_accelerometer} show measurements of the accelerometer on the smartphone and Polar Verity, respectively. The measured activity represented by acceleration is low during the night from 00:00 a.m. to 08:00 a.m., but it increases after 08:00 a.m. for both devices. The smartphone was mostly lying on a table during data collection, thus the activity level is lower compared to that of the Polar Verity Sense. Regarding audio data, we did not find any discontinuities among the individual recordings over the course of 24 hours. For the location, we do not provide visualizations due to privacy reasons. The recorded connectivity data follows the connectivity protocol we defined in Table~\ref{tab:sampling_coverage_connectivity_protocol}.
Even though the network connection was abruptly disabled two times, CLAID was able to handle this and continue sampling. It was able to synchronize the collected data after the connection was re-established. We visualize the incoming data over time in Figure~\ref{fig:results:sampling_coverage_arrival_times}.
 
Figures~\ref{fig:results:sampling_coverage_arrival_times:android} and~\ref{fig:results:sampling_coverage_arrival_times:ios} show the data receiving times of individual samples collected during the 24 hours, normalized to represent the percentage of files already received for each sensor combined in one plot. Figures~\ref{fig:results:sampling_coverage_arrival_times:android_parallel} and~\ref{fig:results:sampling_coverage_arrival_times:ios_parallel} show the same but for each sensor separately. The discontinuities due to interrupted network connection between 10:00 a.m. to 12:00 p.m. and 18:00 to 20:00 p.m. are clearly visible. In our experiment, the synchronization of files using Wi-Fi was faster than synchronization via a cellular connection. This can notably be observed on iOS, where the synchronization via cellular took almost 30 minutes after the connection was reestablished at 12:00 p.m. The synchronization via Wi-Fi only took approximately 10 minutes, both on Android and iOS. Note that the transmission speed may differ between Wi-Fi and Cellular. Therefore, we tested the connection speed at the experiment location before conducting the experiment. On Cellular, we recorded an average download speed of 34.1 Mbit per second and an average upload speed of 26.5 Mbit per second using an LTE connection. On Wi-Fi, we recorded an average download speed of 89.9 Mbit per second and an average upload speed of 29.2 Mbit per second.

\subsection{ML-Model in the Loop}
\label{sec:results:ml_deployment}
Results of our \textit{ML-Model in the Loop} experiment are shown in Table~\ref{tab:results:ml_deployment}. We used the same configuration on Android and iOS. Using the test data sent from the server, we were able to obtain equal results on Android and iOS, i.e., the exact same number of coughs was detected on both devices. The results for the metrics shown in the table, however, only are based on the number of coughs detected. A cough is recognized when the averaged outputs of the models in the ensemble surpass a certain threshold in a number of subsequent windows. Even if the model outputs slightly differ between the devices, a cough is still recognized when the output is above the threshold. We further investigated differences in model outputs directly between the devices, before applying the threshold. Exemplary results for one test file of 6 seconds are shown in Table~\ref{tab:results:ml_deployment_deviations}.
\begin{table}[!htb]

      \centering
      \caption{Results of ML-Model in the Loop Experiment on Android vs. iOS}
        \label{tab:results:ml_deployment}
        \begin{tabular}{lcc}
        \hline
                    & Android      & iOS          \\\hline
        Sensitivity & 0.7518796992 & 0.7518796992 \\
        Specificity & 0.9849498328 & 0.9849498328 \\
        Precision   & 0.9569377990 & 0.9569377990 \\
        Accuracy    & 0.9131944444 & 0.9131944444 \\
        MCC         & 0.7942635659 & 0.7942635659 \\\hline
        \end{tabular}

\end{table}
\begin{table}[!htb]
      \centering
      \caption{Deviations in model outputs}
    \label{tab:results:ml_deployment_deviations}
    \begin{tabular}{lc}
    \hline
                &      Value                      \\\hline
    Number of values & 400                        \\
    Equal values & 181                            \\
    Different values  & 219                       \\
    Max difference    & $6.26\cdot10^{-7}$        \\
    Mean difference         & $6.16\cdot10^{-8}$  \\\hline
    \end{tabular}
\end{table}
For this particular test file, Mel spectrograms were calculated for 40 windows as part of the moving window process. Note, that the number of windows might vary between different files, as explained in Section~\ref{sec:exeriments:ml_evaluation}. Five models are part of the ensemble, each having two output values for cough or non-cough, respectively. This results in $40 \cdot 5 \cdot 2 = 400$ values for this file, representing cough or non-cough probabilities of each model for each window. Of these, 181 are equal and 219 are different between the Android and iOS devices, with max and mean deviations shown in the table. 

\subsection{Battery and memory consumption}
\label{sec:results:battery_memory}
\begin{figure*}[h!t]     \centering
     \begin{subfigure}{0.49\textwidth}
         \centering
         \includegraphics[width=1.0\textwidth]{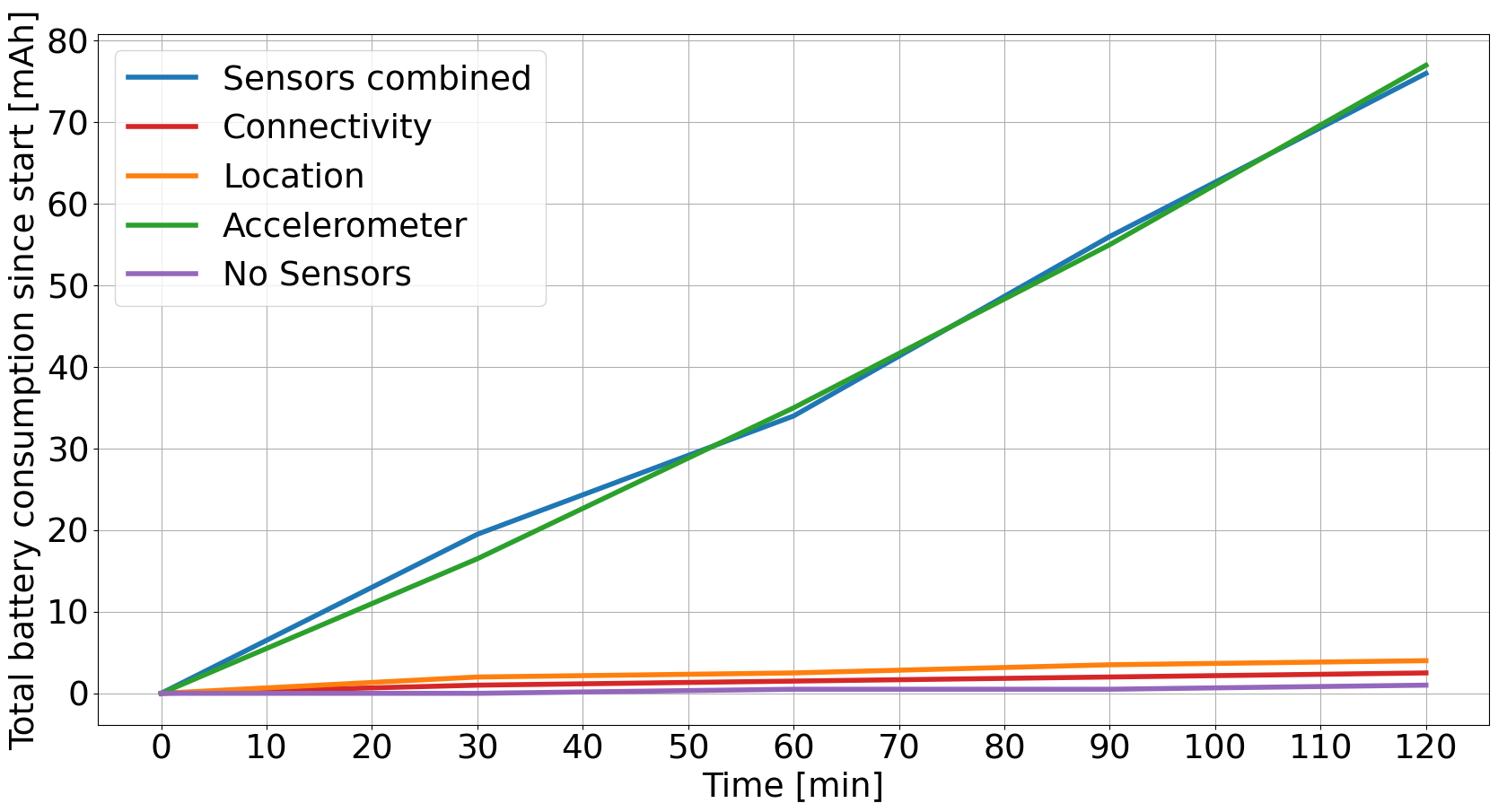}
         \caption{Battery consumption using different sensors~(Android)}
     \end{subfigure}
     \begin{subfigure}{0.49\textwidth}
         \centering
         \includegraphics[width=1.0\textwidth]{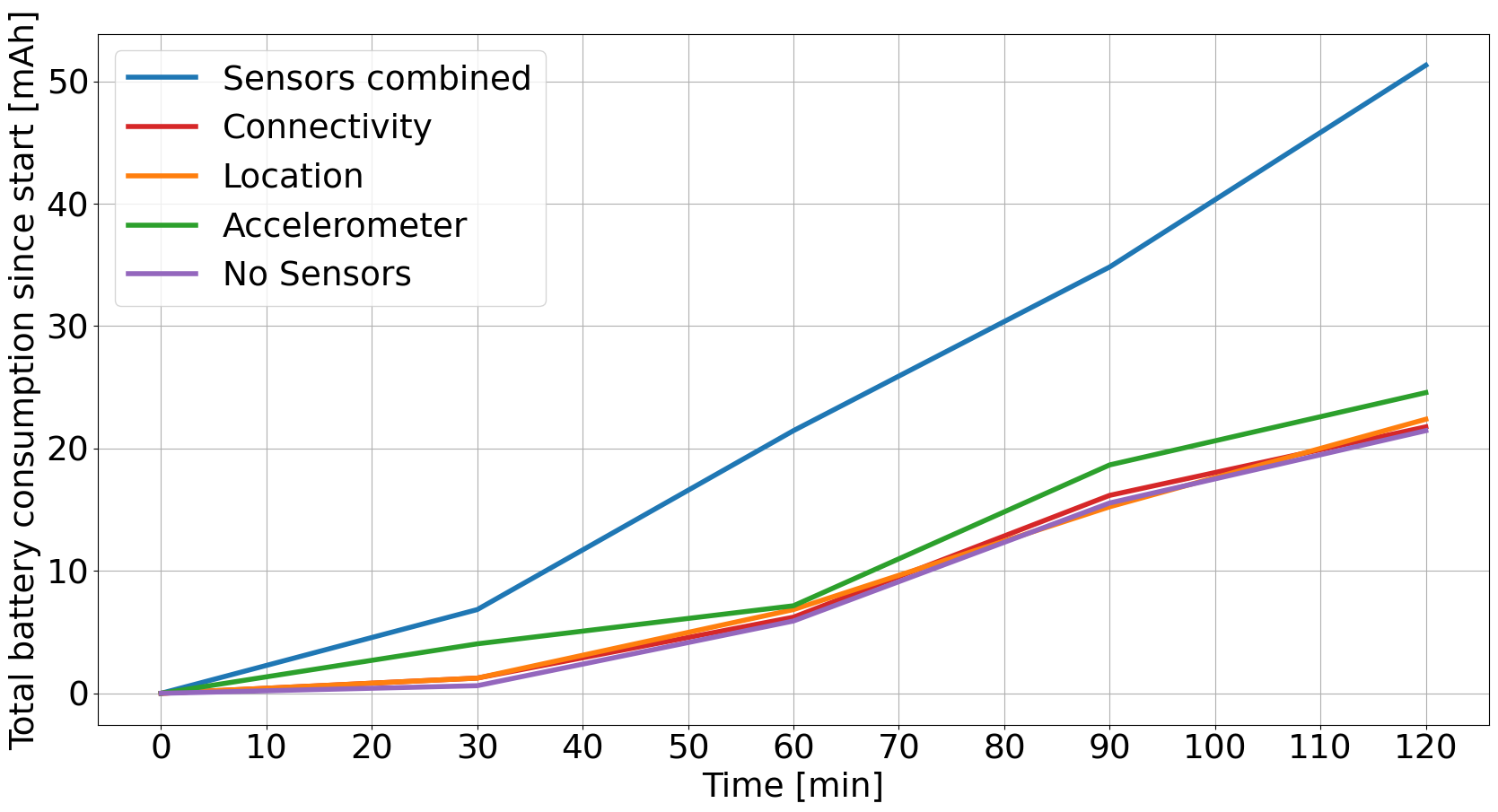}
         \caption{Battery consumption using different sensors~(iOS)}
     \end{subfigure}
    \caption{Battery consumption of CLAID over 2 hours using different sensors. Battery consumption is equivalent to the cumulative draw of battery~(total battery drainage), estimated for our App.}
    \label{fig:results:battery_consumption}

    \centering
     \begin{subfigure}{0.49\textwidth}
         \centering
         \includegraphics[width=1.0\textwidth]{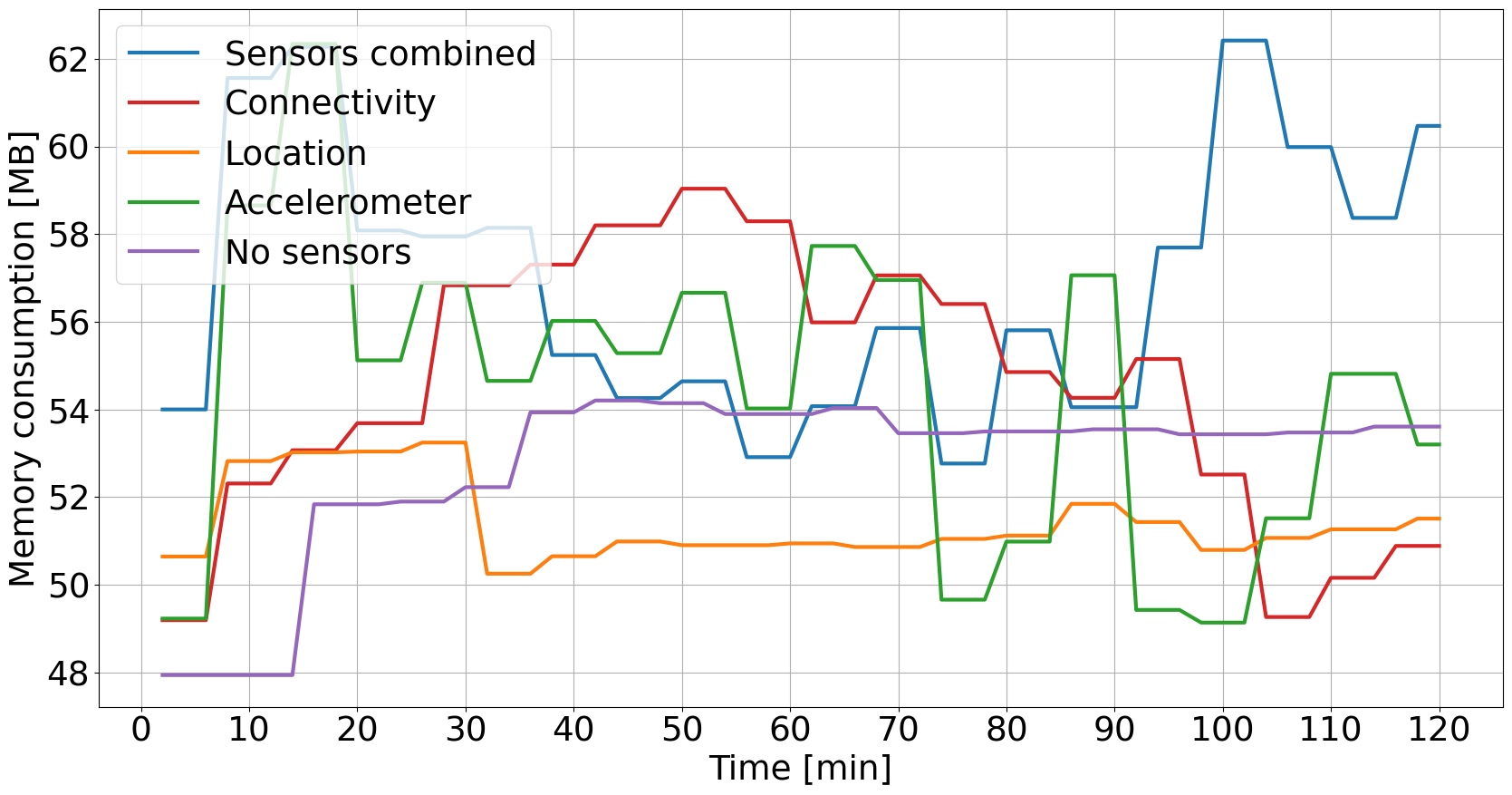}
         \caption{Memory consumption using different sensors~(Android)}
     \end{subfigure}
     \hfill
     \begin{subfigure}{0.49\textwidth}
         \centering
         \includegraphics[width=1.0\textwidth]{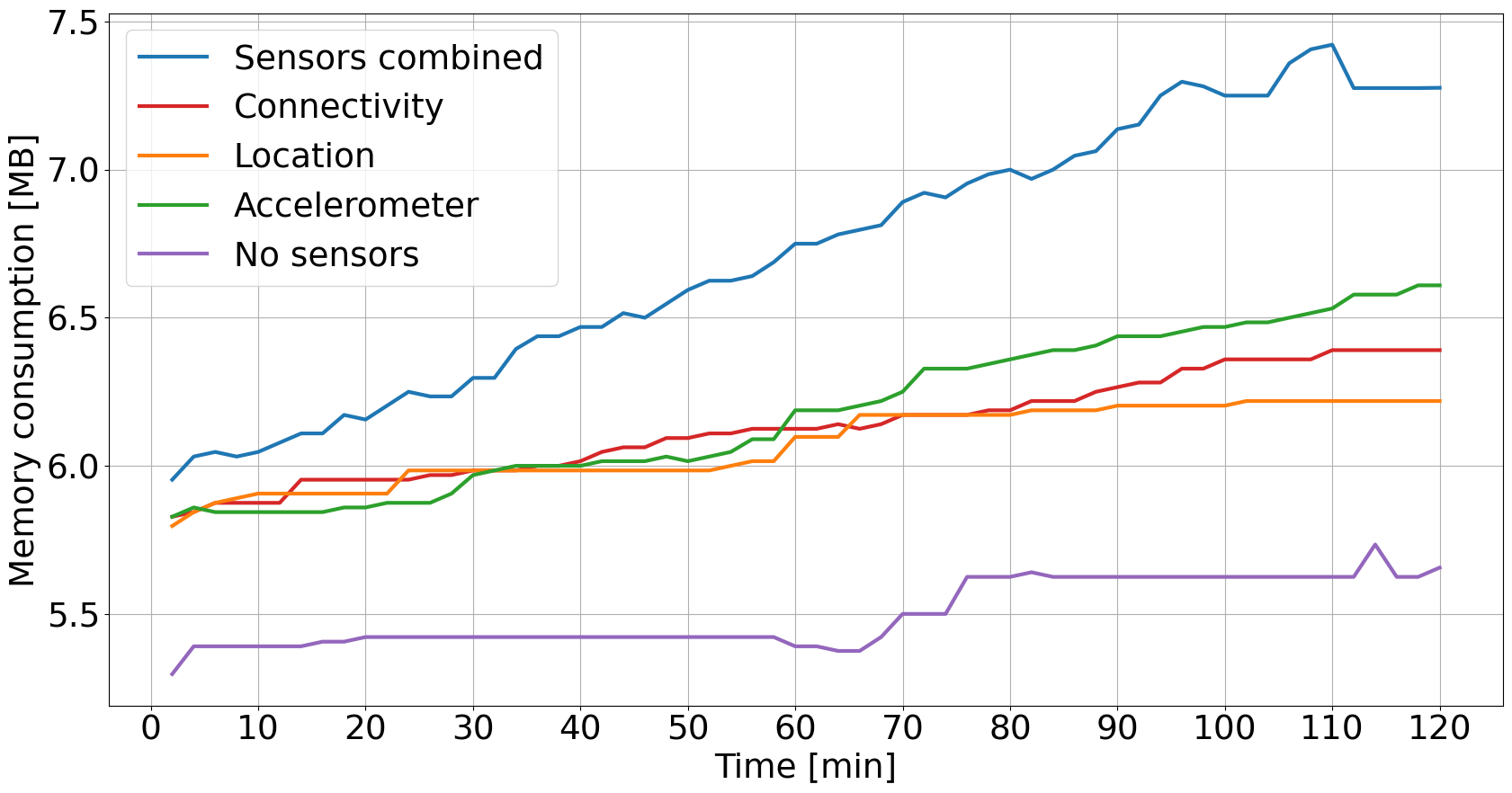}
         \caption{Memory consumption using different sensors~(iOS)}
     \end{subfigure}
     \hfill
    \caption{Memory consumption of CLAID over 2 hours using different sensors. Memory consumption refers to the total amount of RAM consumed by our application on the system, including allocated data as well as space required for the code.}
    \label{fig:results:memory_consumption}
\end{figure*}

We estimated Battery and memory consumption using the Android Studio~(version 2021.2.1) and Xcode~(version 14.2) profilers respectively, which allow estimating consumption for individual applications. Figure~\ref{fig:results:battery_consumption} depicts the battery consumption for both devices.
The starting point for the battery level was 100\% for each sensor configuration. 
On iOS, CLAID consumes more battery compared to Android when idling or using few sensors but is more energy-efficient when utilizing multiple sensors. Both iOS and Android experience the highest battery drain when all sensors are in use, resulting in 76mAh to be consumed on Android and 51.32mAh on iOS, respectively. When using all sensors in combination, battery consumption is mainly determined by the Accelerometer sensor. On Android, connectivity and location sensors consume minimal battery, only drawing 4mAh over two hours, while the accelerometer drains 76mAh in the same time frame. Conversely, on iOS, the battery drain from individual sensors is more balanced, with the accelerometer drawing roughly 24mAh, and the location and connectivity sensors both draining approximately 21mAh over a two-hour period. From the battery consumption, we calculated the current draw of each sensor as shown in Table~\ref{tab:results:battery_consumption:current_draw}.

Analog to the battery level, the memory consumption during the same periods is shown in Figure~\ref{fig:results:memory_consumption}. The Figure shows that the memory consumption for Android is higher than on iOS, with major fluctuations in the consumption profiles. Compared to that, memory consumption on iOS is more stable. It seems to increase during the first hour and then saturates towards stable values. Maximum memory usage is reached when using all sensors in combination on both devices. The maximum memory consumption on Android in that case is 62.42 megabytes~(MB). In contrast, iOS reaches a maximum memory consumption of only 7.50MB when using all sensors in combination. Hence, memory consumption on Android is more than eight times higher in the same scenario. The lowest memory consumption on iOS is achieved when using no sensors, resulting in approximately 5.73MB to be consumed. On Android, memory consumption is still 54.21MB even when using no sensors. Temporarily, using connectivity and location sensors even requires less memory, i.e., 51.50MB and 53.00MB respectively at the end of the 2-hour period. Regarding single sensors, the accelerometer consumes the most memory, with a maximum memory usage of 61.34MB on Android and 6.61MB on iOS. Table~\ref{tab:results:battery_consumption:max_memory_consumption} summarizes the maximum memory consumption of each sensor configuration within the two-hour period.

\begin{table*}[t]

\centering

\begin{tabular}{llllll}
Sensor                             & Combined & Connectivity & Location & Accelerometer & No sensors\\\hline
Current draw [mA] Android & 38.50 & 1.25	& 2.00	& 38.00 &	0.50       \\
Current draw [mA] iOS  & 25.66 & 10.89 & 11.20  & 12.29 & 10.73\\\hline
\end{tabular}
\caption{This table shows the approximate current draw for each sensor, individually as well as combined, calculated from the battery consumption over two hours. For comparison, we also calculated the current draw of CLAID in an idle state while not using any sensors.}
\label{tab:results:battery_consumption:current_draw}
\begin{tabular}{llllll}
Sensor                             & Combined & Connectivity & Location & Accelerometer & No sensors\\\hline
Max. memory usage [MB] Android & 62.42	& 59.04 & 53.25 & 61.34 & 54.21      \\
Max. memory usage [MB] iOS  & 7.50 & 6.44 & 6.23 & 6.61 & 5.73\\\hline
\end{tabular}
\caption{This table shows the maximum memory consumption in terms of RAM usage for each sensor, individually as well as combined, over the two-hour period. For comparison, we also provide the maximum memory consumption of CLAID in an idle state while not using any sensors.}
\label{tab:results:battery_consumption:max_memory_consumption}
\end{table*}
Lastly, we also assessed the memory usage during our 24-hour sampling experiment, using 9 sensors, and found that it reached a maximum of approximately 90MB for Android and 50MB for iOS over the course of 24 hours.

\section{Discussion}
\label{sec:discussion}
In this paper, we introduced and evaluated CLAID, a middleware framework based on transparent computing, which stems from our research in the field of digital biomarkers. It is targeted at addressing the lack of frameworks meeting the requirements for the development of digital biomarkers, specifically modularity~(1), measurements~(2), and verification and validation~(3). Our framework allows the combination of edge and cloud devices into a logical edge-cloud system, unifying capabilities of sensor data collection and deployment of machine learning models across devices. This enables the deployment even of complex digital biomarker modules and allows validation in different scenarios. Principal findings, comparison with prior work, practical implications, and limitations are discussed in the following.

\paragraph{\textbf{Principal findings}}
Our experiments suggest that CLAID meets the requirements for a framework for the development of digital biomarkers in several ways, as reasoned in the following. First, the \textbf{modular} middleware design enables the implementation of flexible and loosely coupled Modules, e.g., for data collection and machine learning, explicitly targeting the modularity requirement of digital biomarkers. It is possible to adapt existing Modules or to add new Modules as required. Our package system allows the integration of new features or extensions to the API. Second, CLAID enables \textbf{measurements} through stable data collection across devices, as demonstrated by Experiment 1. Given that a sensor Module is available, data collection only requires the specification of a configuration file, which is usable across different devices or operating systems. The reuse of configuration files makes it possible to standardize data collection across different devices. The integration of novel sensors only requires the implementation of the corresponding sensor Module. Common pitfalls such as data storage and synchronization are mitigated by our ready-to-use DataSaverModule and DataSyncModule.  Third, our implementation offers different approaches to deploy models for \textbf{verification} and \textbf{validation}. It is possible to integrate existing machine learning pipelines on the server using Python API bindings. Alternatively, our machine learning Modules can load and execute models on the server or directly on the phone. Our approach to transparent computing allows us to combine Modules across different devices. This approach enables the implementation of a novel methodology for the verification of digital biomarker models, which we refer to as ML-Model in the Loop. With experiment 2, we demonstrated how this method allows the transfer of a model from the training environment on the server to a smartphone. After deploying models using this method, they can directly use data of available sensors of the device, simply by adapting the configuration files. 

With Experiment 3, we also evaluated memory and battery consumption in different configurations and found CLAID to be more efficient on iOS than on Android. On Android, memory consumption is approximately 8 times higher, possibly due to overhead introduced by the Java virtual machine and garbage collector. In contrast, iOS memory consumption on iOS is more stable. Regarding battery consumption, Android performs better with fewer sensors, while battery consumption is comparatively lower when using multiple sensors at once. It is important to note that the estimated values for battery consumption are approximate and influenced by factors such as CPU usage, screen-on-time, and overheads introduced by logging utilities in Android Studio and Xcode, which increase battery consumption and induce memory overhead that we are unable to account for.

\paragraph{\textbf{Comparison to prior work}}
Prior work only partially addresses the requirements for the development of digital biomarkers established in Section~\ref{sec:introduction}. For instance, frameworks like Pogo~\cite{pogo_middleware} and USense~\cite{usense} focus on modularity, but have not been developed for medical use cases and do not meet requirements 2 and 3 as they lack support for external sensors and integration of machine learning models. Requirement 2 for measurements is addressed by frameworks such as Sensus, RADAR-BASE, or CAMS~\cite{sensus, radar_base, carp_cams}, which focus on sensing. They have been developed especially for data collection from mobile and wearable devices, partially also for medical use cases. However, they do not provide the required modularity to incorporate machine learning models and therefore do not meet requirements 1 and 3. Lastly, requirement 3 is fulfilled by frameworks such as MobiCOP~\cite{mobicop, mobicop_iot} and MTC~\cite{mtc_new}. While they have not been implemented for medical scenarios, they allow combining edge and cloud resources to offload computations and hence can enable the deployment of machine learning models. However, they do not provide the same flexibility as found in frameworks like Pogo and do not offer support for incorporating measurements by sensors, failing to address requirements 1 and 2.

To the best of our knowledge, there currently does not exist any open-source framework unifying middleware platforms, mobile sensing capabilities, and transparent computing and offloading capabilities, that natively supports the implementation of applications across Android, iOS, and WearOS next to Linux, macOS, and Windows. 

\paragraph{\textbf{Practical implications}}
To facilitate the usage of CLAID, we provide pre-existing packages including Modules for data collection, synchronization, background tasks, and machine learning deployment. Our package manager allows installing and managing these packages. To use these Modules in applications, users can write configuration files, which allow them to specify the names and properties of the required Modules. Based on these files, the framework loads and configures the Modules at runtime. The transparent computing functionality enables seamless data streaming between devices and programming languages. We provide an API that allows the implementation of new and custom Modules. With Python API bindings available on desktop operating systems, it is possible to stream data directly from a phone to Python programs on a server. This capability enables live visualizations and data analysis, for example via packages for feature generation from wearable data like FLIRT~\cite{flirt}.
This approach also enables the deployment of machine learning models on the server by integrating existing Python code. Alternatively, we provide machine learning Modules to load models from a file in a supported format, for example using TensorFlow Lite, which can either be used on the server or the phone as required. These Modules can load machine learning models from the file system, the assets~(Android), bundle~(iOS), or from a network path. 
 
\paragraph{\textbf{Limitations}}
We consider the following points to be limitations of CLAID, based on our experiments and the implementation of the framework:
\begin{enumerate}
    \item \textit{Long-term evaluation}: While we have demonstrated the capabilities of our framework in experiments, we have to further evaluate it in our ongoing studies to assess reliability and effectiveness over longer periods of time and in different studies. Additionally, we did not evaluate our system in terms of memory and battery consumption compared to other frameworks. 
    \item \textit{Encryption}: Data privacy and security are major concerns when collecting data from patients during medical studies. Currently, CLAID permits the transmission of pseudonymized data using hashes or non-personal user identifiers, however, does not yet provide support for encryption during data storage or transmission. We are currently developing a prospective solution that leverages our reflection system to integrate encryption support for various data types.
    \item \textit{Sustainability}: To handle the complexity of multiple devices, operating systems, and sensors, we require a systematic approach for the continuation of the framework, involving documentation and automated tests. While we provide preliminary documentation on our website, we do not yet include automated component or integration tests when updating the framework. We are currently working on a strategy for the long-term sustainability of our platform.
   
\end{enumerate}

\section{Conclusions and Outlook}
\label{sec:outlook}
To the best of our knowledge, CLAID contributes to the field of digital biomarkers by (1) offering a modular open-source middleware framework, serving as a platform to build digital biomarkers (2) enabling stable data collection across various devices, and (3) providing methods to deploy trained machine learning models for verification and validation. The provided API allows to implement Modules in different programming languages such as C++, Java, or Python. Existing packages offer Modules that enable data collection from various sensors on smartphones or wearables, as well as the deployment of machine learning models. Using principles of transparent computing, CLAID allows combining applications running on devices with any of the major operating systems, namely Linux, macOS, Windows, Android, iOS, and WearOS. Modules written in any supported language can communicate across devices in order to exchange data or offload tasks.

While we presented common scenarios involving data collection and deployment of machine learning models in the research area of digital biomarkers, our system, however, is not limited to the use cases presented in this work. Since CLAID Modules are loosely coupled, it is possible to adapt or replace them to suit new use cases or integrate new Modules as required. The extension is facilitated by the use of a package system that allows the integration of new functionalities. Since multiple instances of CLAID can be connected, it is possible to realize different network topologies as required, e.g., connecting multiple smartphones or other devices. There is no fixed hierarchy between connected devices, and arbitrary topologies are possible. 

In our future work, we will further evaluate our framework in the field. Currently, we are conducting a study to gather data from participants using multiple sensors, including a Samsung Galaxy Watch running CLAID on WearOS. Another planned study will involve various sensors like smartwatches, spirometry devices, and audio data, accompanied by surveys and interactive elements. These studies aim to contribute to the development and evaluation of novel digital biomarkers with the support of machine learning. We will test these digital biomarkers in real-world scenarios using our platform's model deployment capabilities. We aim to enhance compatibility with more machine learning algorithms and enable the deployment of different and larger models. To achieve this, we intend to integrate more machine learning Modules, such as PyTorchMobile and CoreML, and optimize preprocessing Modules for improved efficiency, enabling support for a broader range of machine learning frameworks. Additionally, we are currently implementing support for encryption for data storage and upload. We are also working on improving the long-term sustainability of our project by providing further documentation and setting up automated test environments.

\appendix
\section*{Acknowledgement}
We would like to thank Francesco Feher for his implementations and contributions to the evaluation of our system.

 \bibliographystyle{elsarticle-num} 
 \bibliography{main}

\end{document}